\documentclass[usenatbib]{tjaa} 
\usepackage[utf8]{inputenc}
\newsavebox\verbbox
\usepackage{float,placeins}

\title[Discovery of a pulsating star with IST40 telescope]{Discovery of a Short Period Pulsator from Istanbul University Observatory}

\author[M. Turan Sağlam et al.]{
M. Turan Sağlam$^{1,2}$\orcid{0000-0001-8903-3770}\thanks{E-mail: mustafaturansaglam@ogr.iu.edu.tr},
Meryem Çördük$^{2}$\orcid{0000 0002 1145 5890},
Sinan Aliş$^{1,2}$\orcid{0000-0002-6990-8899},
Görkem Özgül$^{2}$\orcid{0000-0002-7291-5571},
\newauthor{
Olcaytuğ Özgüllü$^{2}$\orcid{0000-0002-4785-683X},
Fatih Erkam Göktürk$^{2}$\orcid{0000-0003-2528-6183},
Rahmi Gündüz$^{1,2}$\orcid{0000-0002-2706-7841},
Süleyman Fişek$^{1,2}$\orcid{0000-0002-3187-5286},}
\newauthor{
F. Korhan Yelkenci$^{1,2}$\orcid{0000-0003-2675-3564},
E. Kaan Ülgen$^{2}\orcid{0000-0002-8056-4214}$,
Tolga Güver$^{1,2}\orcid{0000-0002-3531-9842}$
}
\\
$^{1}$Istanbul University Observatory Research and Application Center, 34116 Istanbul, Turkey\\
$^{2}$Department of Astronomy and Space Sciences, Faculty of Science, Istanbul University, 34116 Istanbul, Turkey
}

\date{Accepted: XXX. Revised: YYY. Received: ZZZ.}
\renewcommand{\pubyear}{0000}
\renewcommand{\volume}{0}
\renewcommand{\issue}{0}

\begin{document}
\label{firstpage}
\pagerange{\pageref{firstpage}--\pageref{lastpage}}
\maketitle{M00-0000}

\begin{abstract}
We report the discovery of a new short period pulsating variable in the field of exoplanet host star XO-2. Variable has been identified while it was being used as a comparison star. In order to verify the variability of the candidate, a follow-up program was carried out. Period analysis of multi-band light curves revealed a very prominent and consistent pulsation periodicity of $P\sim0.95$ hours. Given the variability period, amplitude and the color index, the object is most likely a \emph{Delta Scuti} type variable. Absolute magnitude ($M_{v}$) and the color index $(B-V)_{0}$ of the star determined as $2.76$ and $0.22$, respectively. This $(B-V)_{0}$ of the star corresponds to A7 spectral type with an approximate effective temperature of 7725 K. Machine-learning analysis of the time-series data also revealed that the object is of variable type DSCT with a probability of 78\%.
\end{abstract}

\section{Introduction}

Delta Scuti stars are pulsating variables with spectral types between A and F, especially from late A toward early F. They are known with their radial and non-radial pulsations spanning periods from 20 minutes to 8 hours \citep{Breger2012}. These variables are mostly located on the zero-age main-sequence in Hertzsprung-Russell diagrams within the so-called instability strip. The amplitude of their variability mostly very small such as $\Delta V = 0.01-0.03$ mag. A subclass of Delta Scuti stars is called as HADS (High-Amplitude Delta Scuti stars) due to their high-amplitude variations ($\Delta V = 0.3-1.0$ mag) in their light curves. Both classes are more massive ($1.2 - 2.4 M_{\odot}$) and hence more luminous than Sun \citep{McNamara11}.

Pulsation behaviour of Delta Scuti stars is an important tool to understand the stellar interiors where Hydogen shell burning takes place. These stars show both multi-mode pulsations with radial or non-radial directions. These modes or pulsation direction can exhibit different behaviours even for the same star \citep{Mow2016}. Studying the changes in these states may provide insights about stellar structures. Therefore, dedicated observing campaigns are carried out both photometric \citep{Breger1993,Breger1994,Zhou2002} and spectroscopic \citep{Ventura2007}.

Recent studies showed that many primary components in Algol and beta-Lyrae type eclipsing binaries are in fact Delta Scuti type stars \citep{KahramanAlicavus2017,Ulas2020}. Investigating these variables in binaries present an important advantage for determining the stellar masses accurately which then can be used to model the stellar interiors.

Robotic telescopes or surveys from space allow to detect numerous Delta Scuti variables like in the case of TESS \citep{Antoci2019} or OGLE \citep{Soszynski2021}. All these stars are then used to construct or update the period-luminosity relation which is a powerful tool to derive luminosities and thus distances for Delta Scuti variables \citep{McNamara11,Poro2021}.

In this study, we report the discovery of a new Delta Scuti variable and present a comprehensive CCD photometry of the object. The new variable has counterparts in both SDSS and PanSTARRS which we provide the appropriate magnitudes in Table \ref{infotab}. The star also observed with GAIA and has a parallax in the EDR3 release. We determined the fundamental pulsation frequency and the ephemeris of the star for the first time. In Section 2, we briefly summarize the observations and data reduction, Section 3 presents light curves and period analysis of the star, Section 4 discusses the variability type and the physical parameters of the star and we summarize our study in Section 5.



\section{Observations and Data Reduction}
\subsection{CCD Observations}
Exoplanet host star XO-2 has been observed within a research program\footnote{The project, 118F042, is supported by the 1001 program of TUBITAK.} that investigates the transit timing variations with a similar fashion given in \cite{Basturk2022}. During these observations, variability of one of the comparison star in the field was noticed.

The field was observed by the 0.4m Schmidt-Cassegrain telescope (aka. IST40) of the Istanbul University Observatory. Observations were carried out with a thermoelectric cooled CCD consisting a KAF-8300 chip which has \emph{3358x2536} pixels. Pixel size of 5.4$\mu$ yields 0.27"/pixel resolution at the focal plane and this resolution allows to capture \emph{16x12} arcminutes field of view.

CCD observations were performed over 10 nights between 23 February 2020 and 27 November 2020 at Istanbul University Observatory. Log of observations is given in Table \ref{obslog}.

\begin{table*}
\caption{Log of observations.}
\begin{center}
\renewcommand{\arraystretch}{1.4}
\setlength\tabcolsep{3pt}
\begin{tabular}{cccccccc}
\hline
Date         & JD Interval				& Duration	&  Number  	 & Filter	& Exposure Time \\ 
                  &   2458000+  				& (hours)		&	of Frames	&		 & (seconds)	     \\
\hline
23.02.2020 & 903.3274 $-$ 903.4571 & 	3.11 & 88 & R & 120 \\
25.02.2020 & 905.2154 $-$ 905.4392 &   5.37 & 231 & Clear &  60 \\
29.02.2020 & 909.3988 $-$ 909.4941 &   2.28 & 122   & Clear & 60  \\
01.03.2020 & 910.2290 $-$ 910.4517 &   5.34 & 200    & Clear & 90  \\
02.03.2020 & 911.2562 $-$ 911.5427 &   6.87 & 150  & R &  75 \\
02.03.2020 & 911.2552 $-$ 911.5418 &   6.87 & 150  & V &  75 \\
03.03.2020 & 912.2345 $-$ 912.5142 &   6.71 & 81  & R &  120 \\
03.03.2020 & 912.2330 $-$ 912.5127 &   6.71 & 82  & V &  120 \\
08.03.2020 & 917.3634 $-$ 917.4376 &   1.78 & 32  & R &  60 \\
08.03.2020 & 917.3626 $-$ 917.4368 &   1.78 & 32  & V &  60 \\
08.03.2020 & 917.3618 $-$ 917.4360 &   1.78 & 33  & B &  60 \\
12.03.2020 & 921.2732 $-$ 921.4006 &   3.05 & 82  & R &  60 \\
12.03.2020 & 921.2724 $-$ 921.3998 &   3.05 & 82  & V &  60 \\
17.11.2020 & 171.3382 $-$ 171.5486 &  5.05 & 64  & B &  150  \\
27.11.2020 & 181.3060 $-$ 181.6867 &  6.73 & 62  & B & 150 \\
\hline
\end{tabular}
\end{center}
\label{obslog}
\end{table*}

All frames were bias, dark and flat-field corrected in a standard manner. Several bias and dark frames were co-added in order to create a master combined calibration frame. Flat-fielding was done with sky flats obtained at dusk. Calibration images were obtained in each observing night.

\subsection{Photometry}

Following the identification of the variable, a comprehensive follow-up program was carried out in February, March, and November 2020 with a total of 10 nights. After confirmation of the candidate as a new variable star using unfiltered observations, standard Johnson-Cousins B, V, and R filtered frames were also taken.

\begin{figure}
\centering
  \includegraphics[width=3in, height=3in]{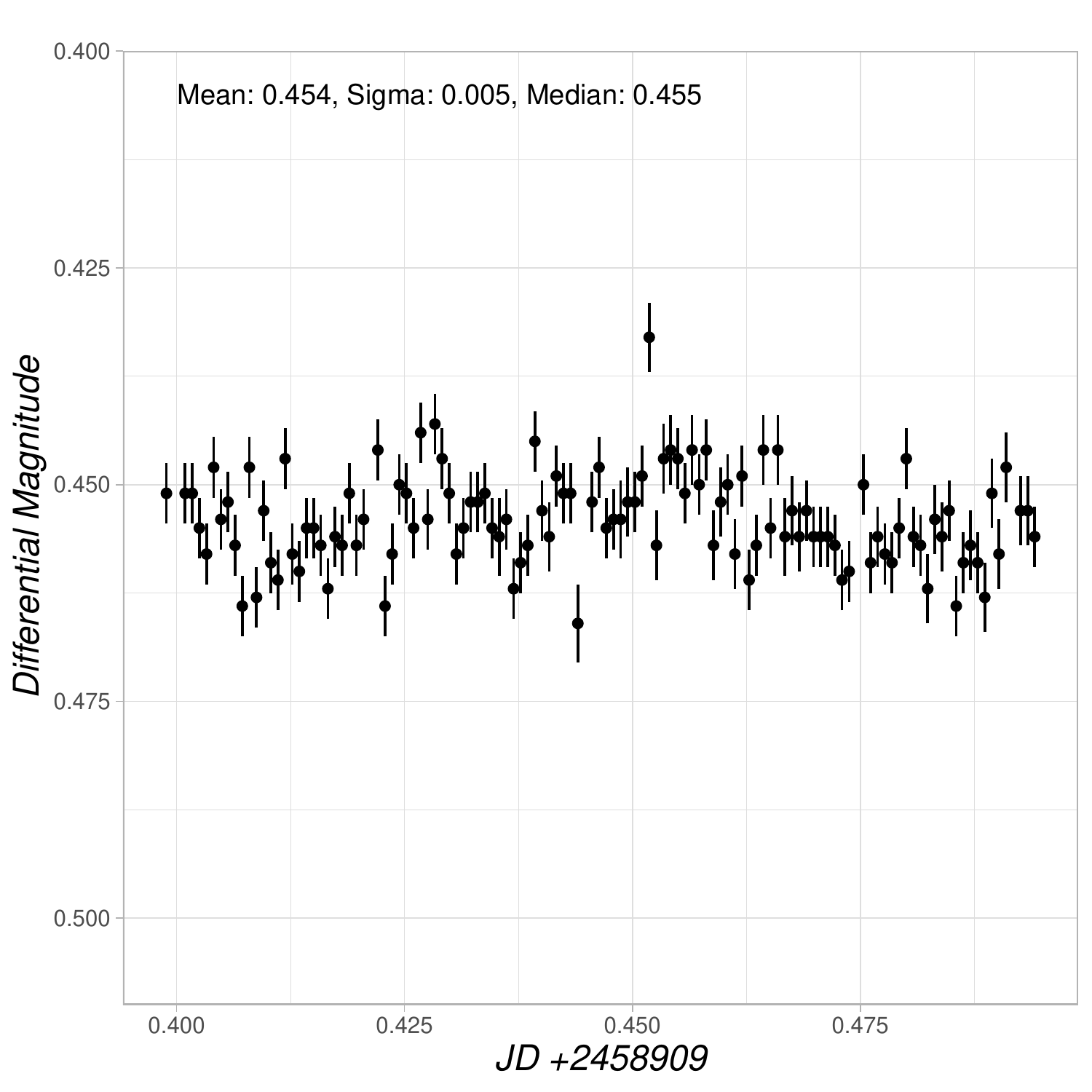}
  \caption{Light curve of C1-C2 difference obtained on 29 February 2020. Frame-to-frame variations of 5 mmag ensure that both comparison stars are not variable.}
\label{c1c2}
\end{figure}

In order to determine comparison stars a thorough inspection of the field stars, almost a dozen, was undertaken. After several field stars examined and differential light curves constructed, stars denoted as C1 and C2 were determined as the most reliable comparison stars. A representative magnitude variation between C1 and C2 is given in Figure \ref{c1c2}. Typical magnitude variation among observing nights varies between 5 mmag and 9 mmag. Strong reliability of frame-to-frame variations ensure neither C1 nor C2 is a variable star, at least in the duration of the present observations.

\begin{figure}
\centering
\includegraphics[width=2.8in, height=1.9in]{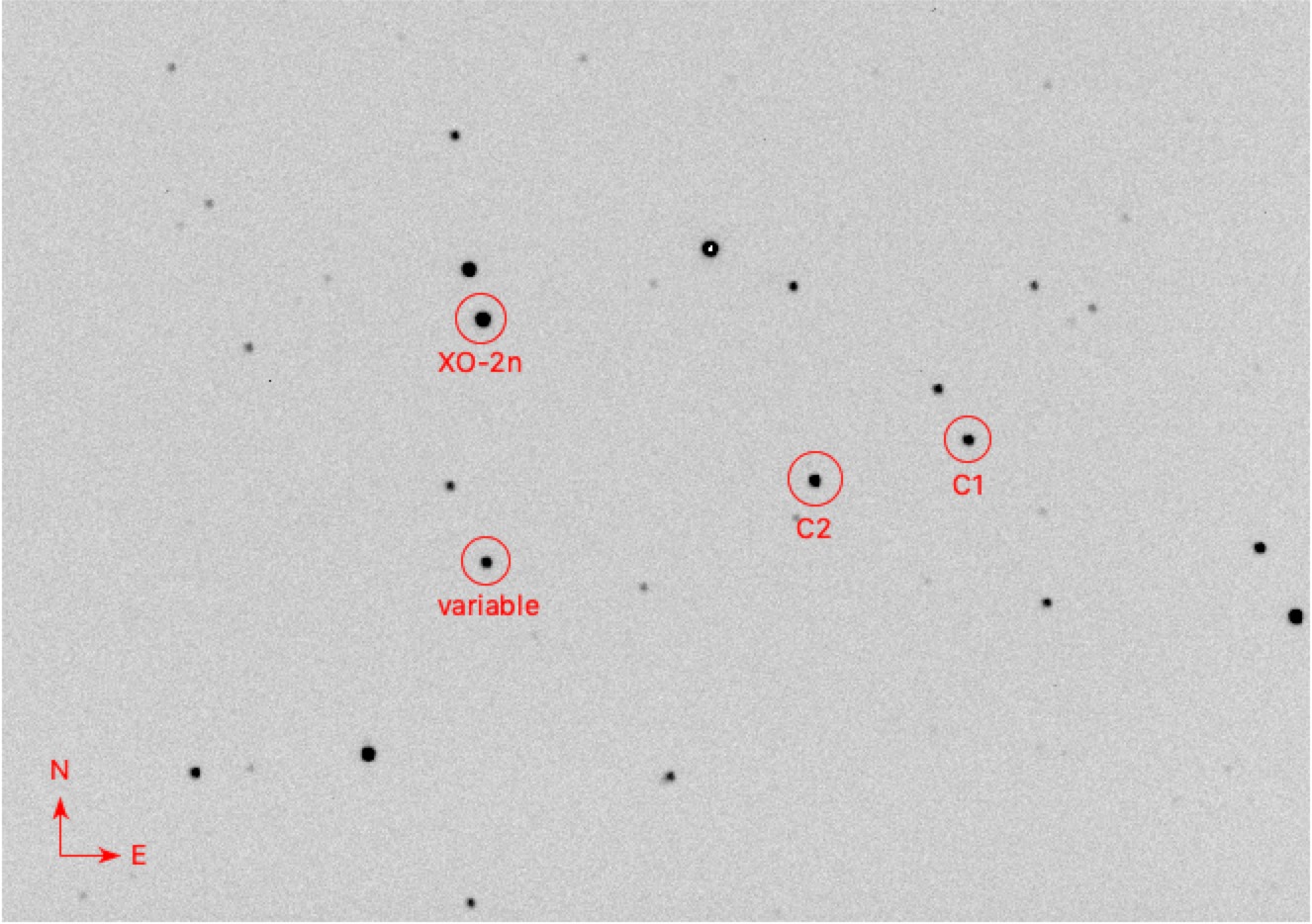}
\caption{Field of the variable taken from IST40 telescope on the night 23 February 2020. Variable and comparison stars are marked as well as the exoplanet host star XO-2.}
\label{varfield}
\end{figure}

Coordinates and basic information of the newly identified variable, comparison stars and the original target of exoplanet host star, XO-2, are given in Table \ref{infotab} and those stars are marked in the field image given in Figure \ref{varfield}.

Instrumental magnitudes were determined with aperture photometry using \emph{Muniwin} software of the \emph{C-Munipack}\footnote{http://c-munipack.sourceforge.net} package \citep{2014ascl.soft02006H}. Photometry procedures of the \emph{C-Munipack} package are based on the well-known DAOPHOT \citep{1987PASP...99..191S} package.

Since both comparison stars, C1 and C2, show constant magnitudes over the observation period, differential magnitudes presented in this study were computed using the comparison star C2.

\begin{table*}
\caption{Basic information of the variable and comparison stars.}
\begin{center}
\renewcommand{\arraystretch}{1.4}
\setlength\tabcolsep{3pt}
\begin{tabular}{cccccccccc}
\hline
Star & RA  & Dec  & u & g & r & i & z & Remark \\ 
     & (J2000) (deg) & (J2000) (deg) & (mag) & (mag) & (mag) & (mag) & (mag) \\ 
\hline
Variable & 117.022937  & 50.265486   & 15.009 & 12.849 & 12.791 & 12.856 & 13.638  & SDSS J074805.50+501555.7\\
C1 & 116.898869  & 50.241248  & 14.857 & 13.139 & 12.690  & 12.587 & 13.160 & SDSS J074735.72+501428.4 \\
C2 & 116.938290  & 50.249308  & 14.882 & 12.651 & 12.212 & 12.067 & 13.319 & SDSS J074745.18+501457.5 \\
XO-2 & 117.026923    & 50.225643 & 14.921 & 14.798 & 14.491  & 10.797 & 13.095  & SDSS J074806.46+501332.3 \\
\hline
\end{tabular}
\label{infotab}
\end{center}
\end{table*}

\section{Light Curve Analysis}

\begin{figure*}
\centering
  \includegraphics[width=4in, height=3in]{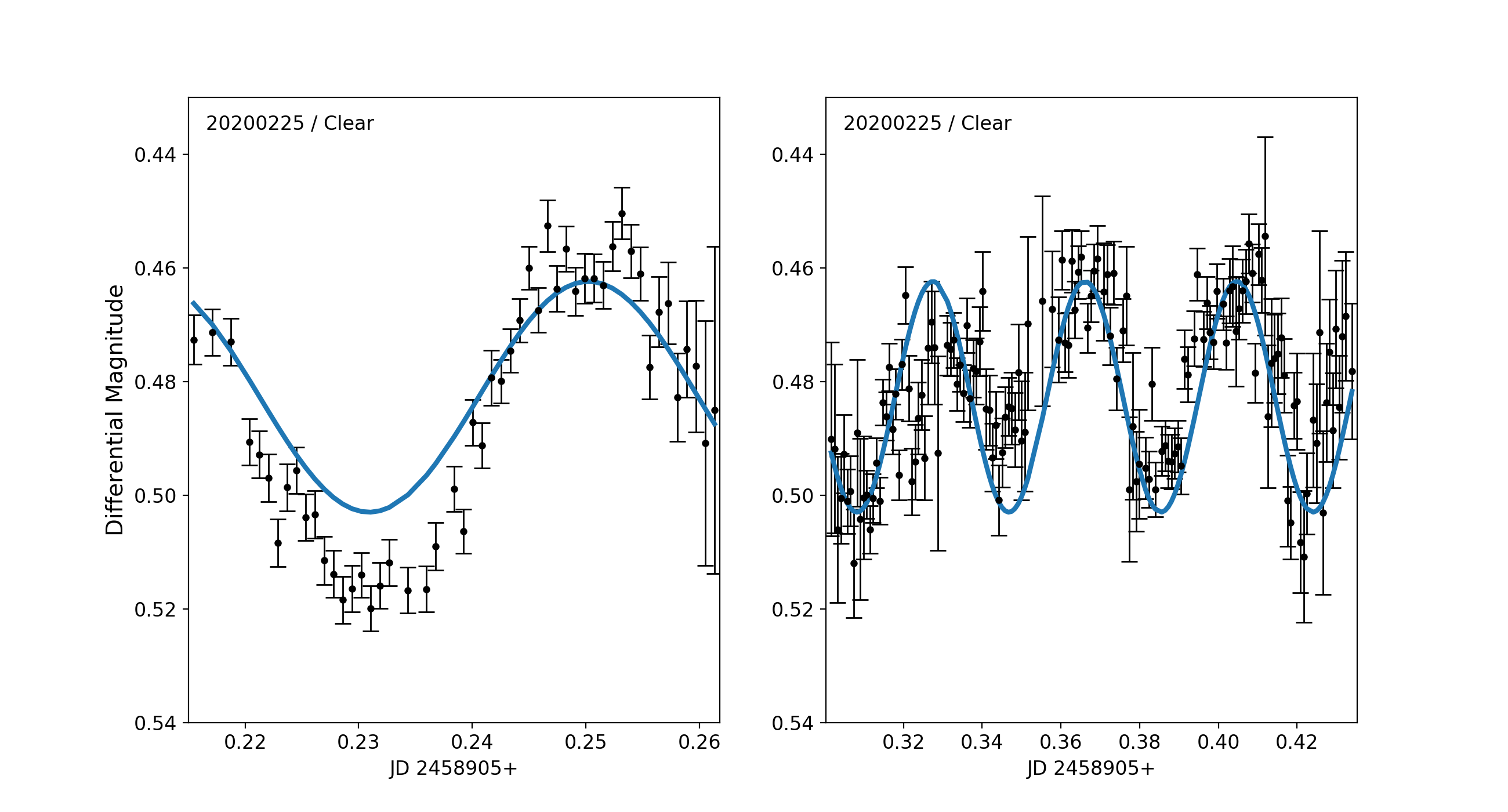}
  \includegraphics[width=3in, height=3in]{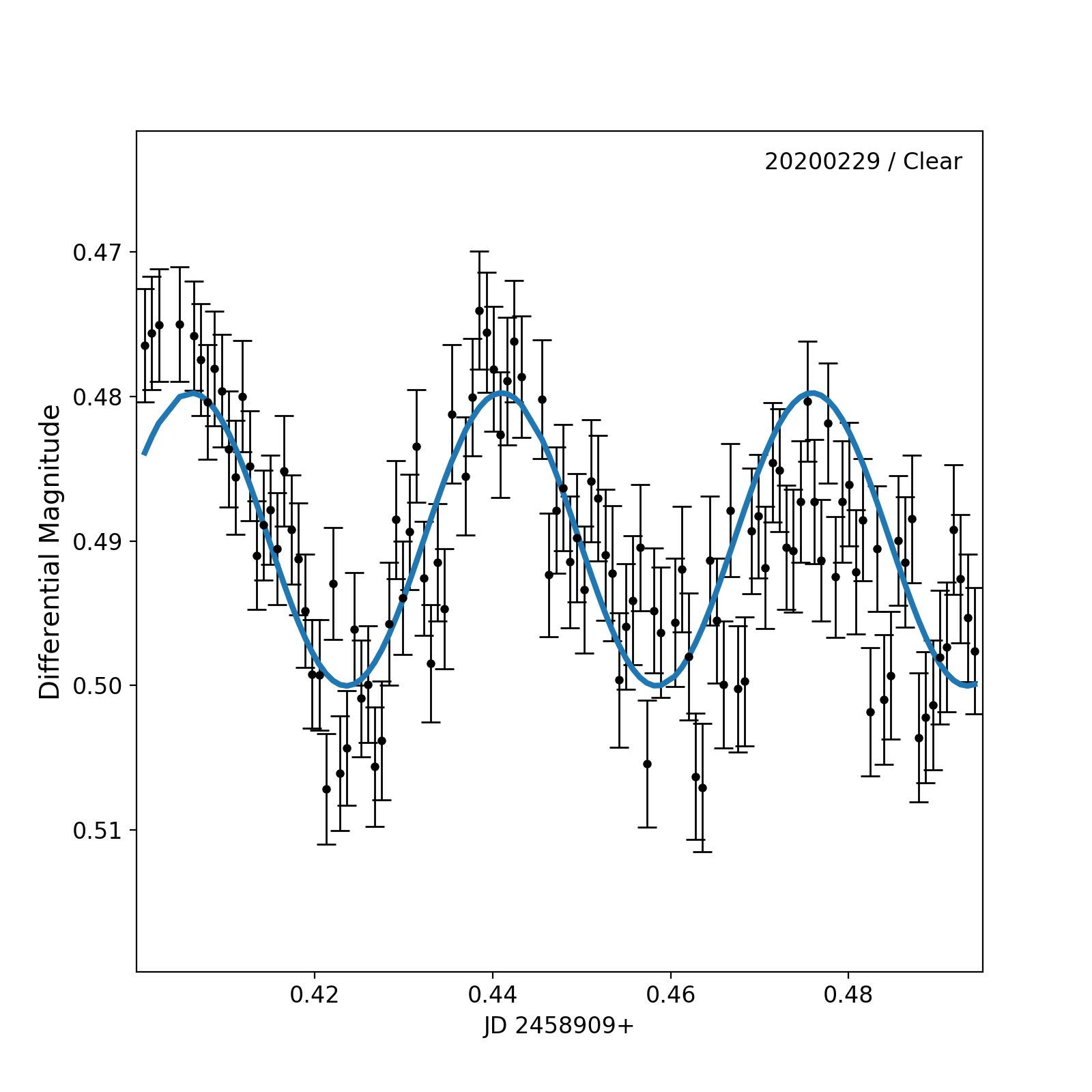}
  \includegraphics[width=3in, height=3in]{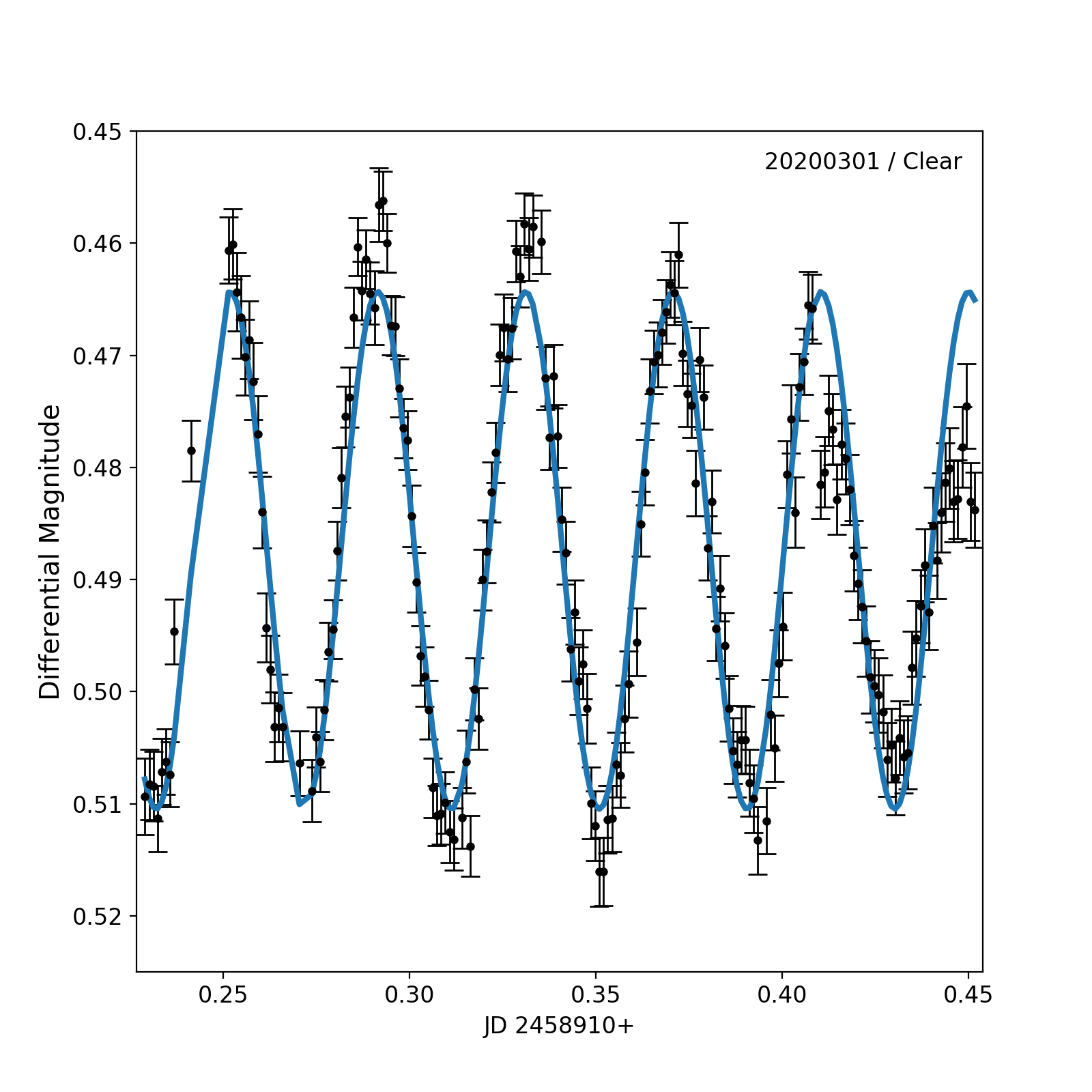}
  \caption{Unfiltered (clear) light curves of the candidate variable. Model curves (blue lines) obtained from the period analysis are also plotted. Details of the maxima times and periods are given in Table \ref{tabfreq} and \ref{tabmaxima}.}
\label{lc1}
\end{figure*}

\begin{figure*}
\centering
  \includegraphics[width=3in, height=3in]{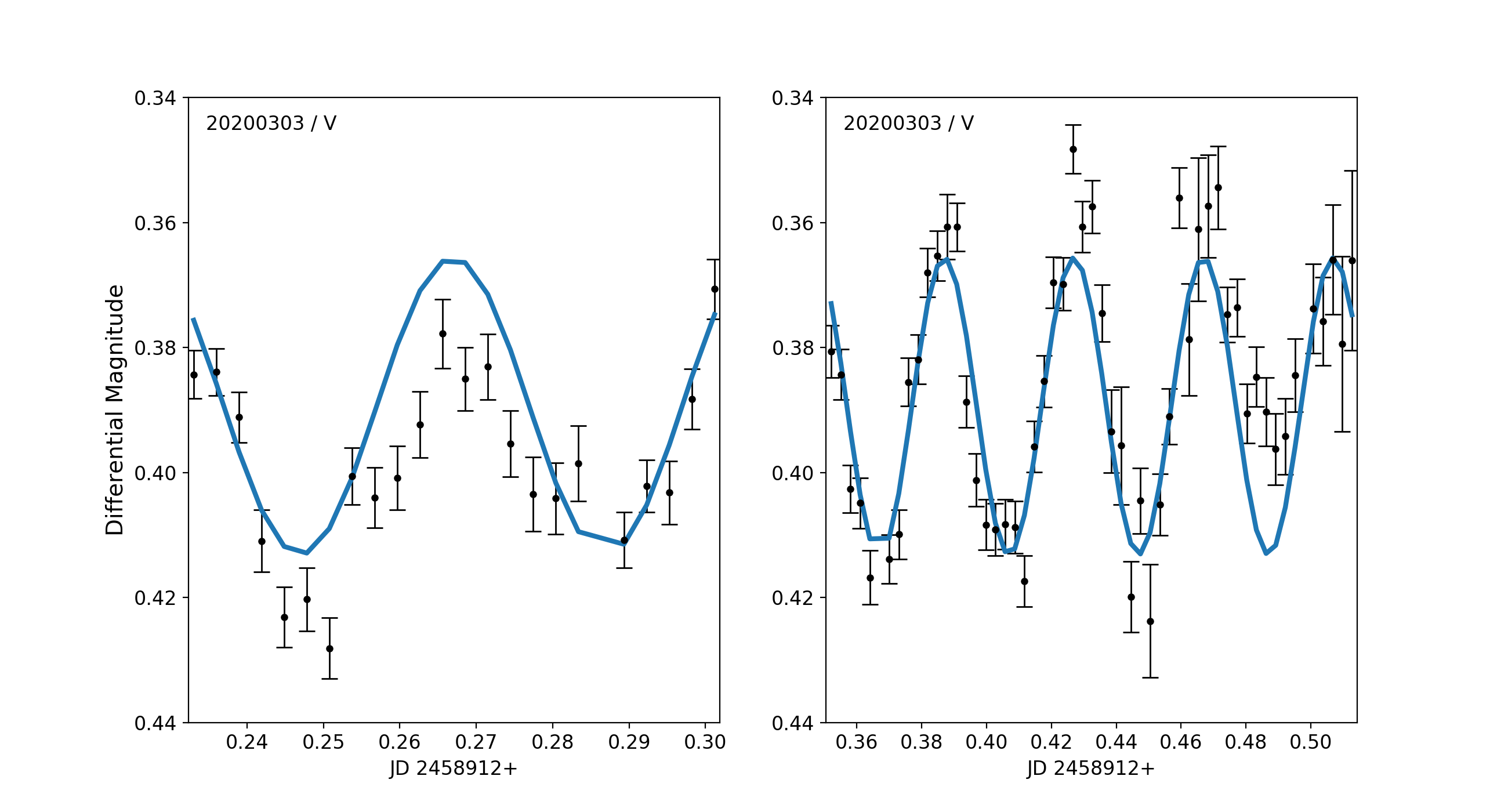}
    \includegraphics[width=3in, height=3in]{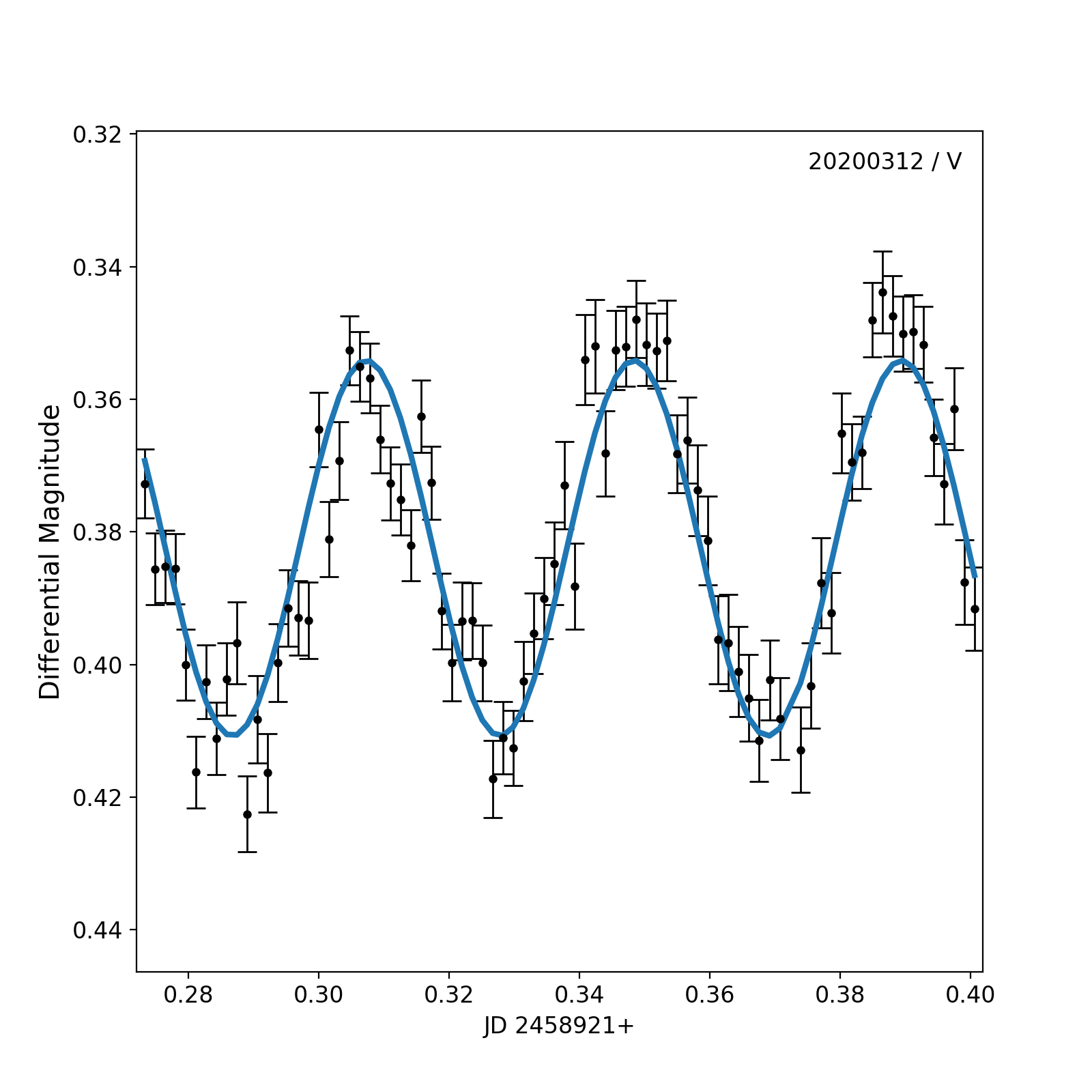}
      \includegraphics[width=3in, height=3in]{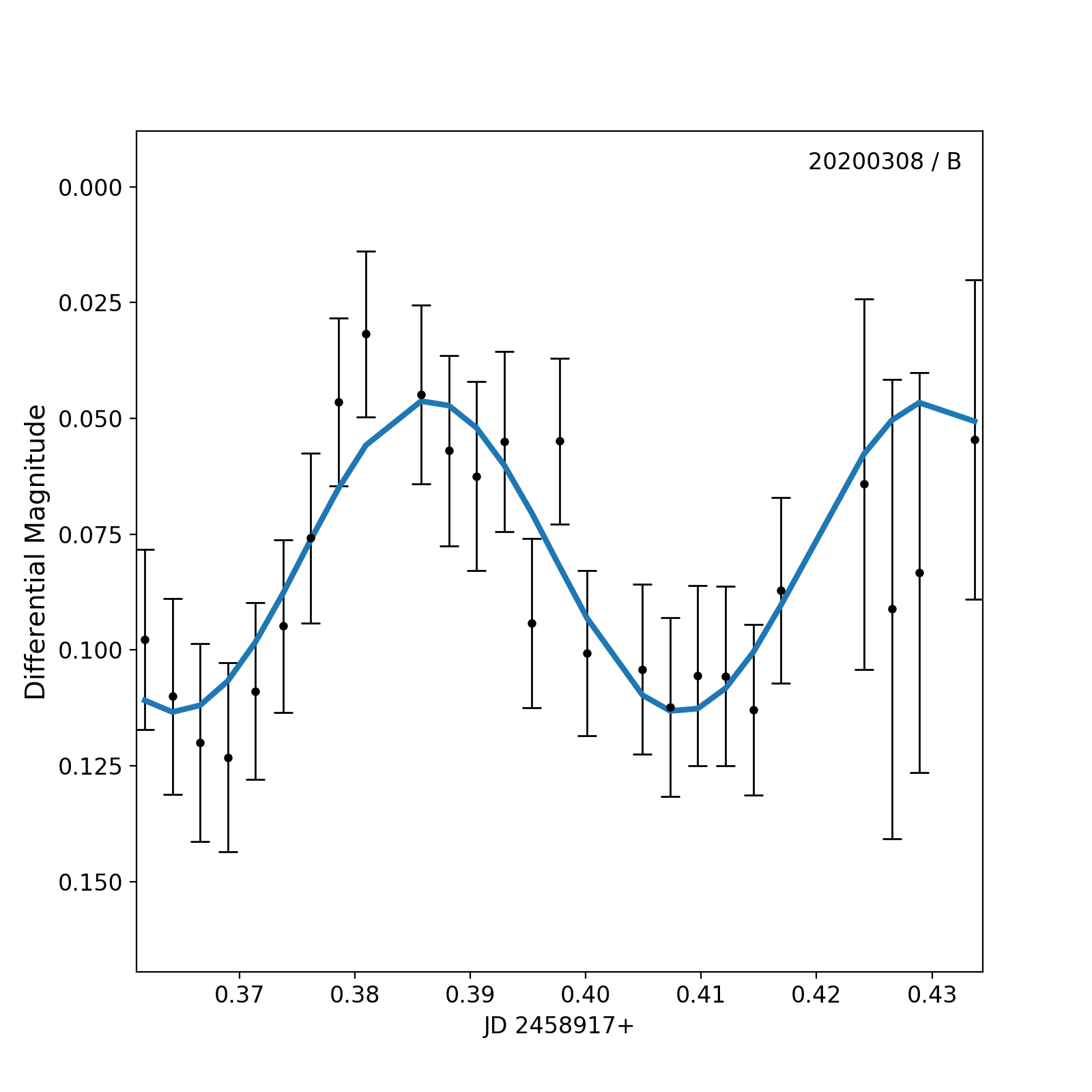}
  \includegraphics[width=3in, height=3in]{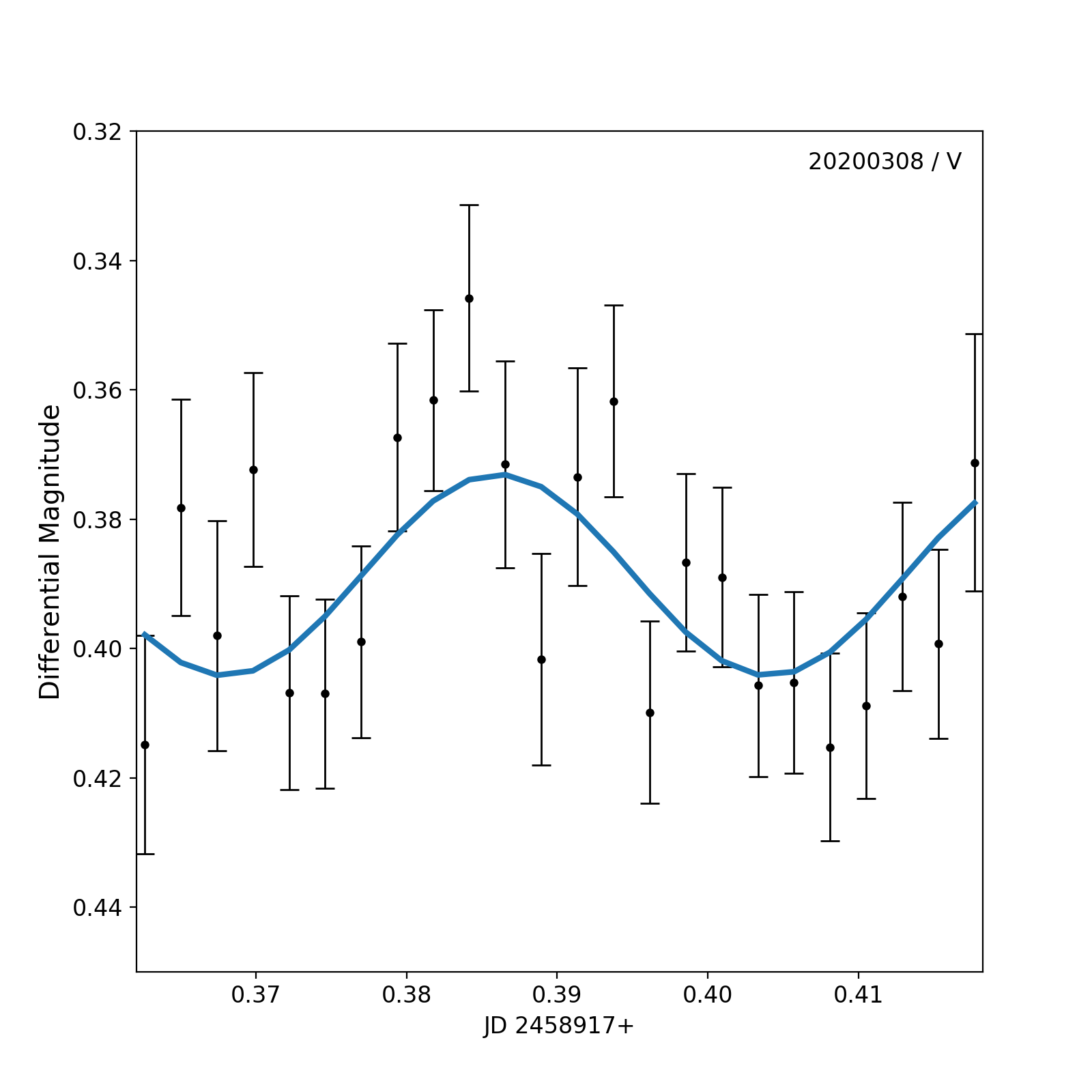}
  \includegraphics[width=3in, height=3in]{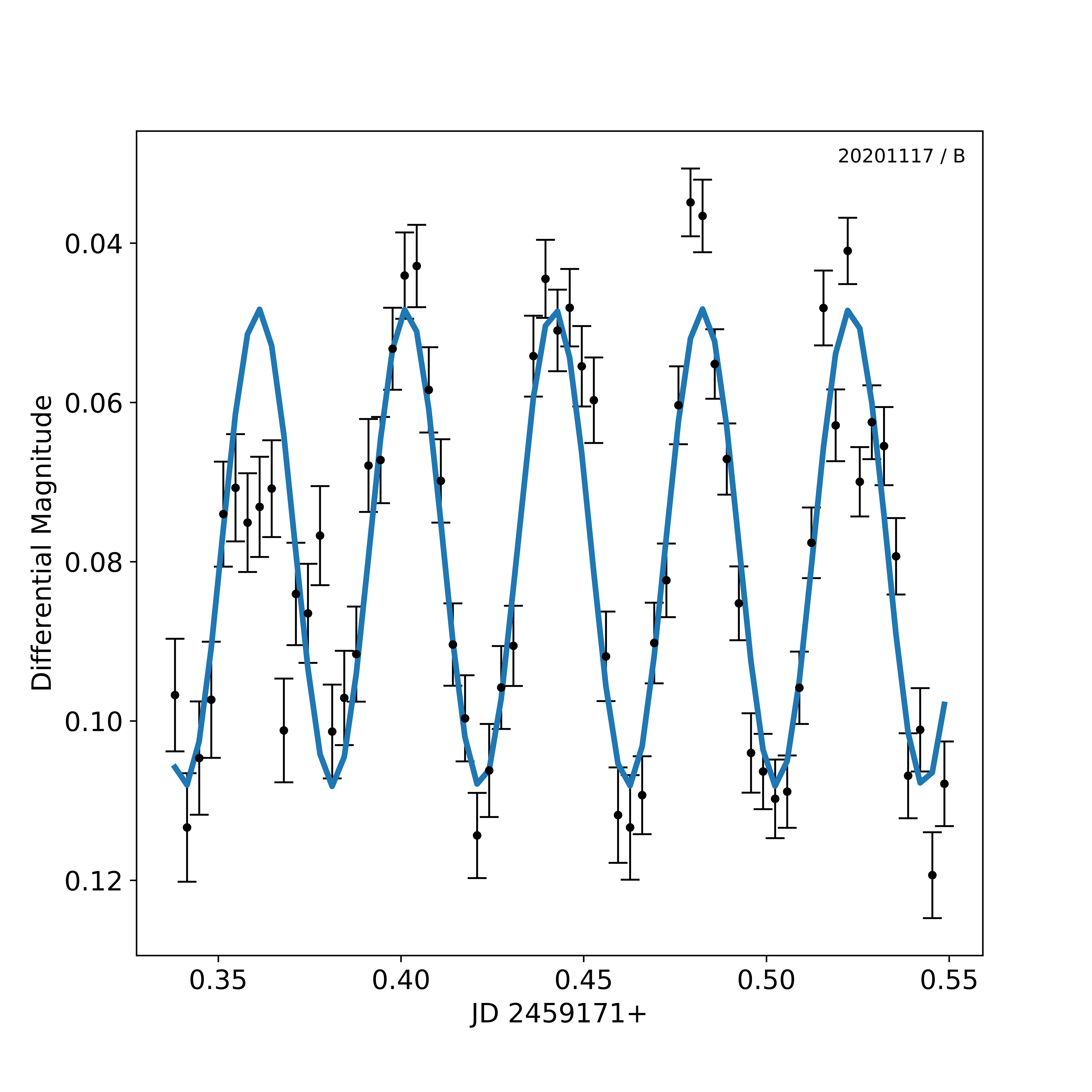}
  \includegraphics[width=3in, height=3in]{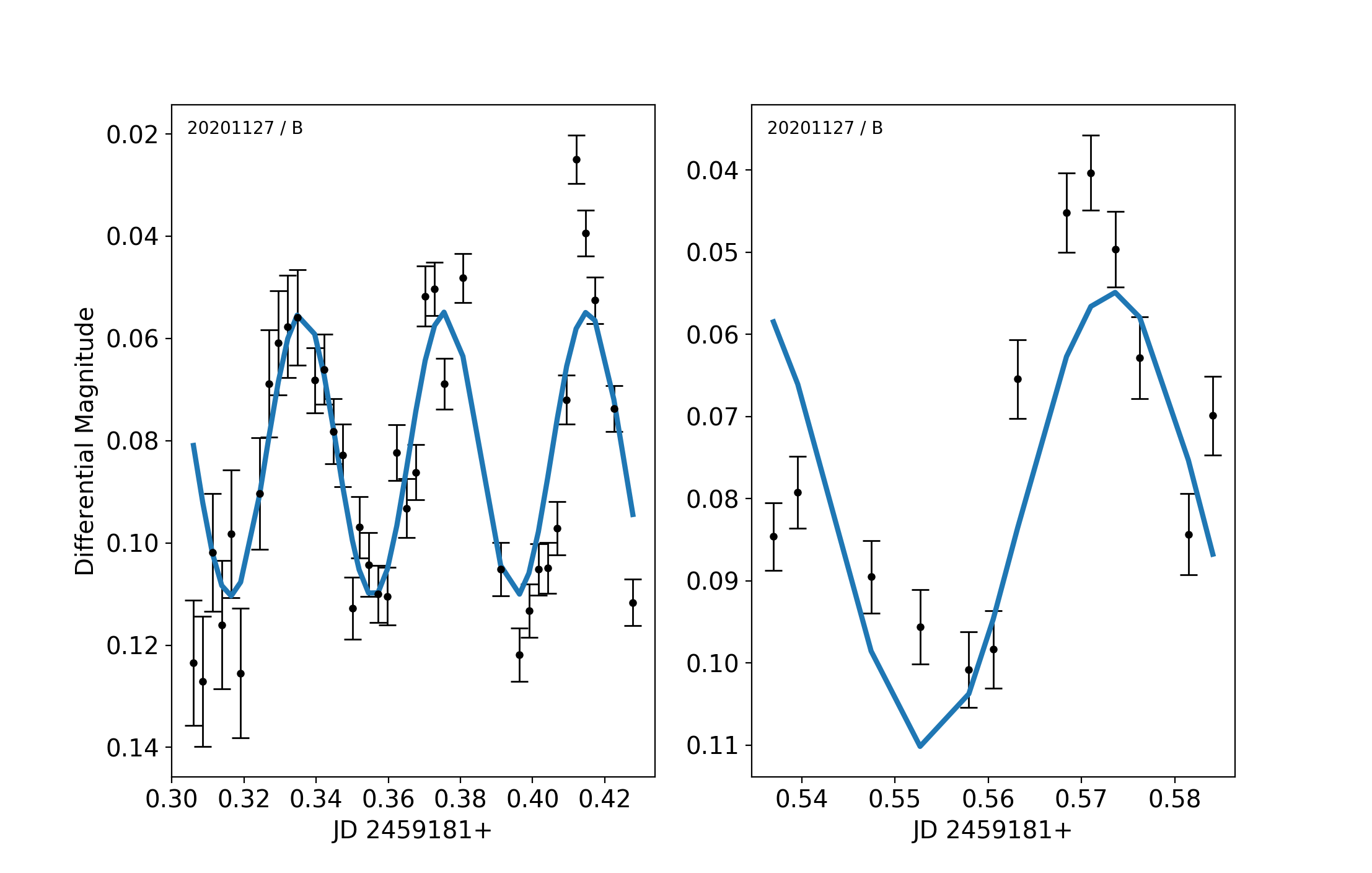}

  \caption{B and V-band light curves of the candidate variable. Model curves (blue lines) obtained from the period analysis are also plotted. Details of the maxima times and periods are given in Table \ref{tabfreq} and \ref{tabmaxima}.}
\label{lc2}
\end{figure*}

\begin{figure*}
\centering
  \includegraphics[width=3in, height=3in]{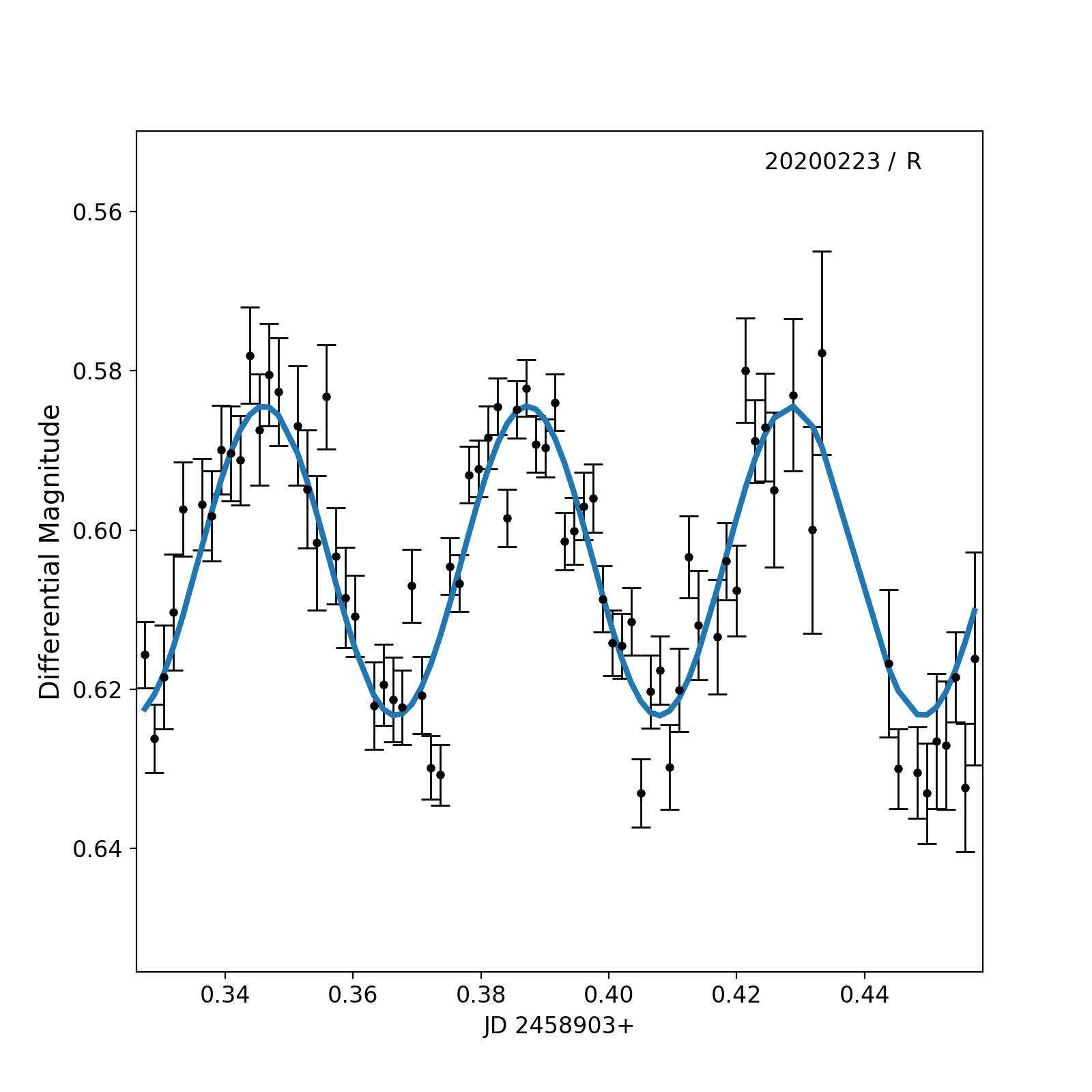}
  \includegraphics[width=3in, height=3in]{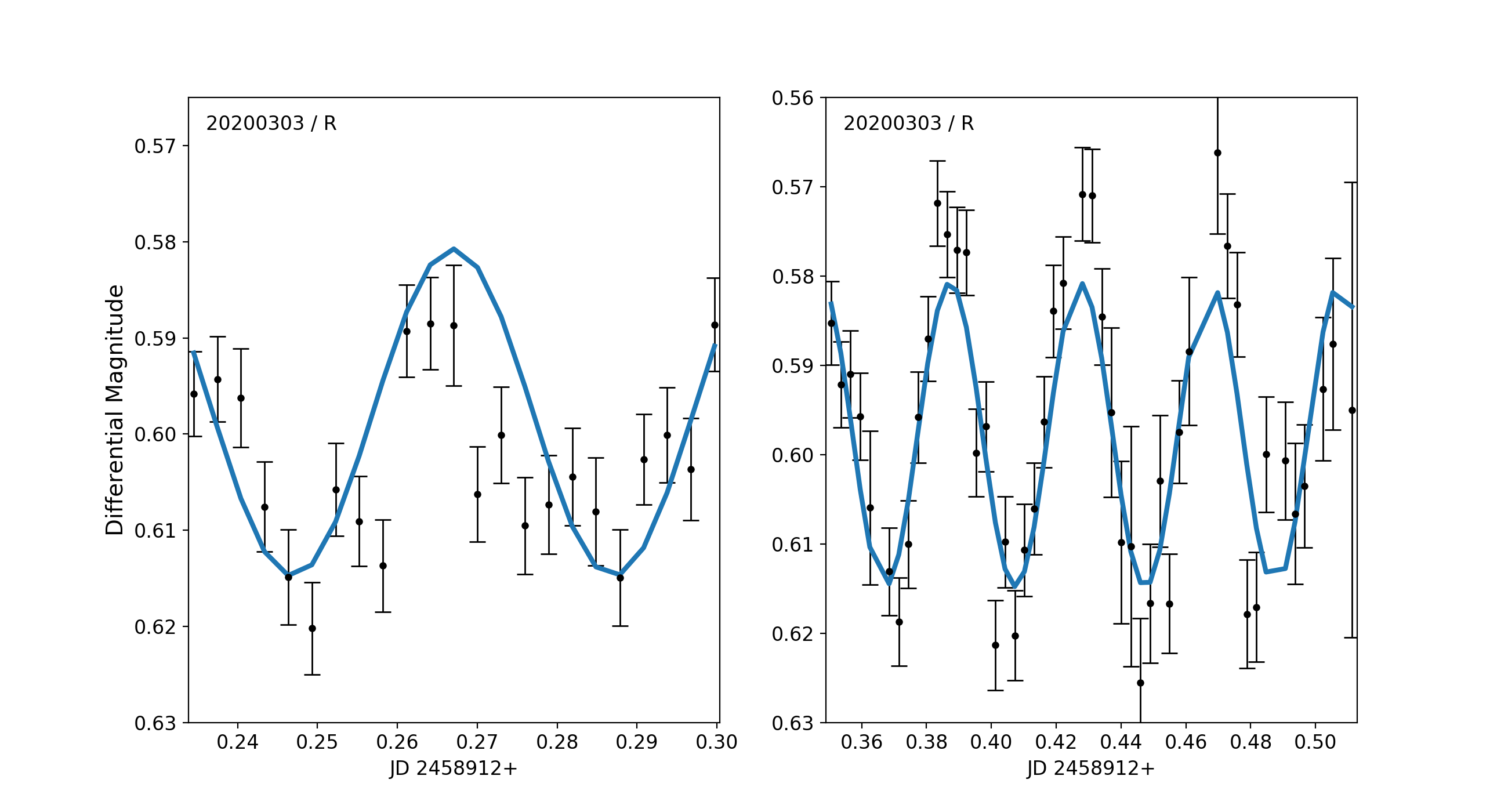}
  \includegraphics[width=3in, height=3in]{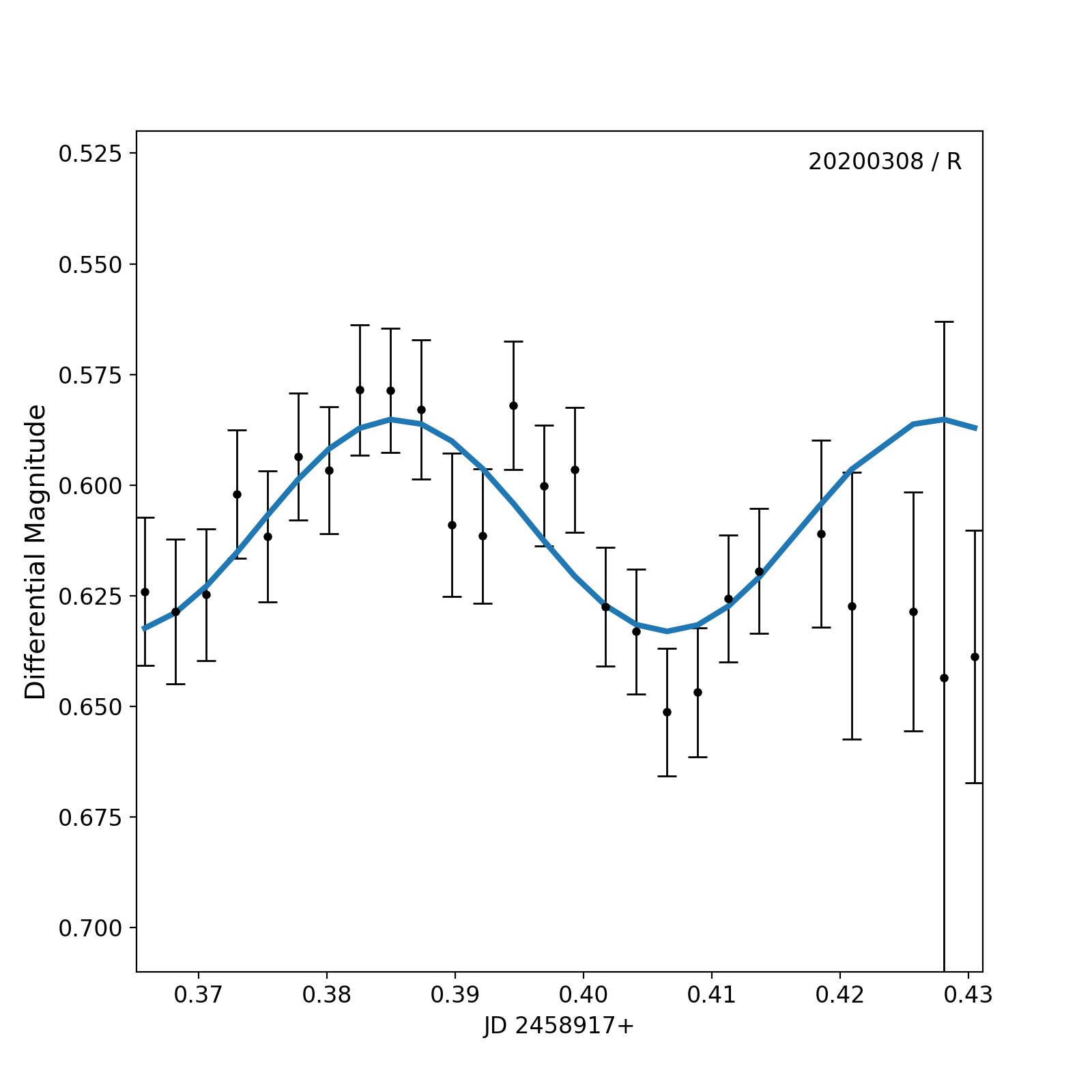}
  \includegraphics[width=3in, height=3in]{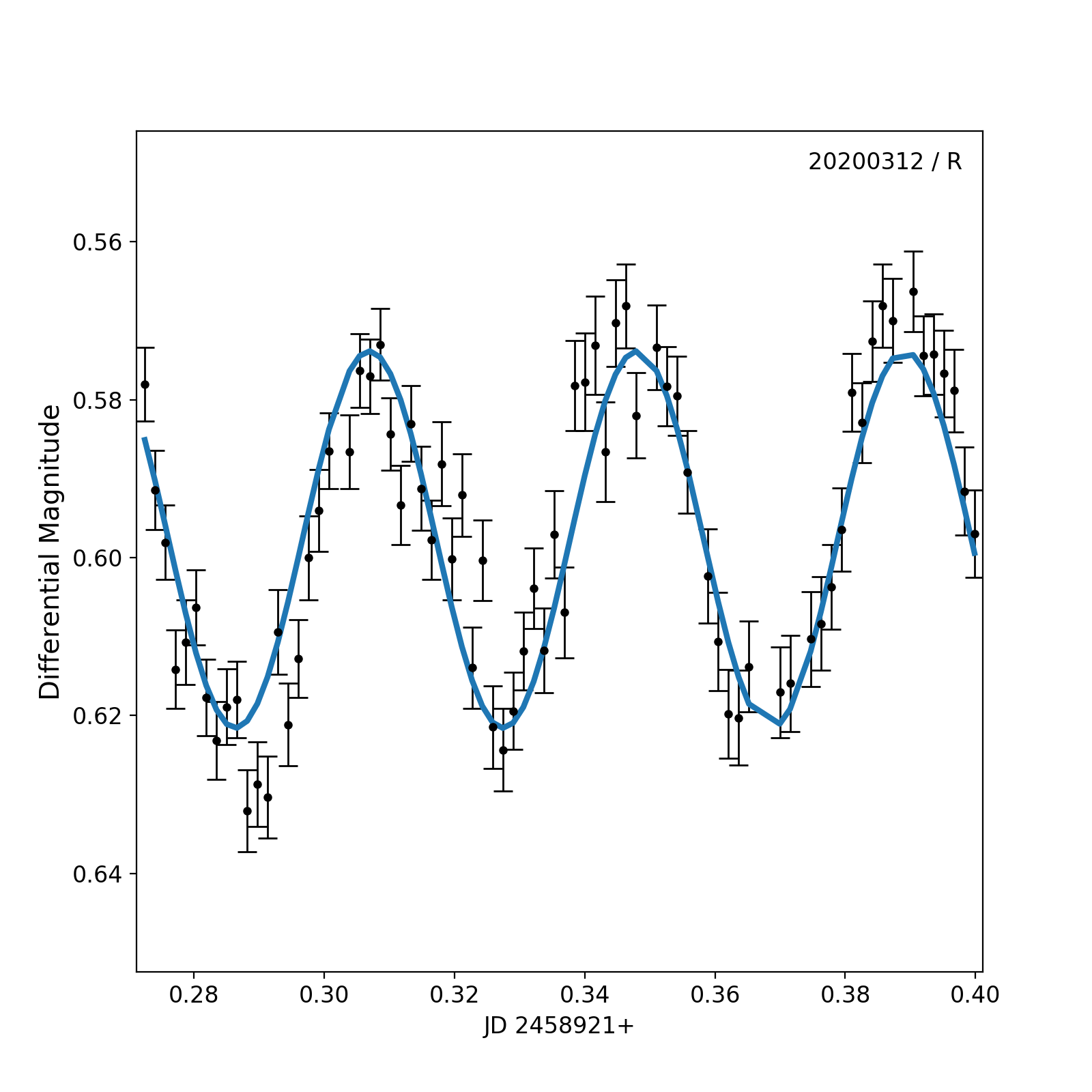}
  \caption{R-band light curves of the candidate variable. Model curves (blue lines) obtained from the period analysis are also plotted. Details of the maxima times and periods are given in Table \ref{tabfreq} and \ref{tabmaxima}.}
\label{lc3}
\end{figure*}


\subsection{Period Analysis}

In order to obtain period of the variation we make use of Period04 \citep{LenzBreger05} which is extensively used in the asteroseismology community. Period04 allows us to model variations by means of Fourier modes and it is possible to include multi-frequency variations in the final model. However, our aim is to characterize the newly discovered variable, hence we model the variations with a single frequency to avoid the complexity. Moreover, the discontinuity in our observations and different filters used in our study would not allow a thorough multi-frequency analysis. Additionally, it is worth noting that the variable has also been observed by the Transiting Exoplanet Survey Satellite (TESS) at a long (i.e. 30 minutes) cadence. The available fluxes are not suitable for a detailed analysis.

\begin{figure*}
\centering
  \includegraphics[width=3in, height=3in]{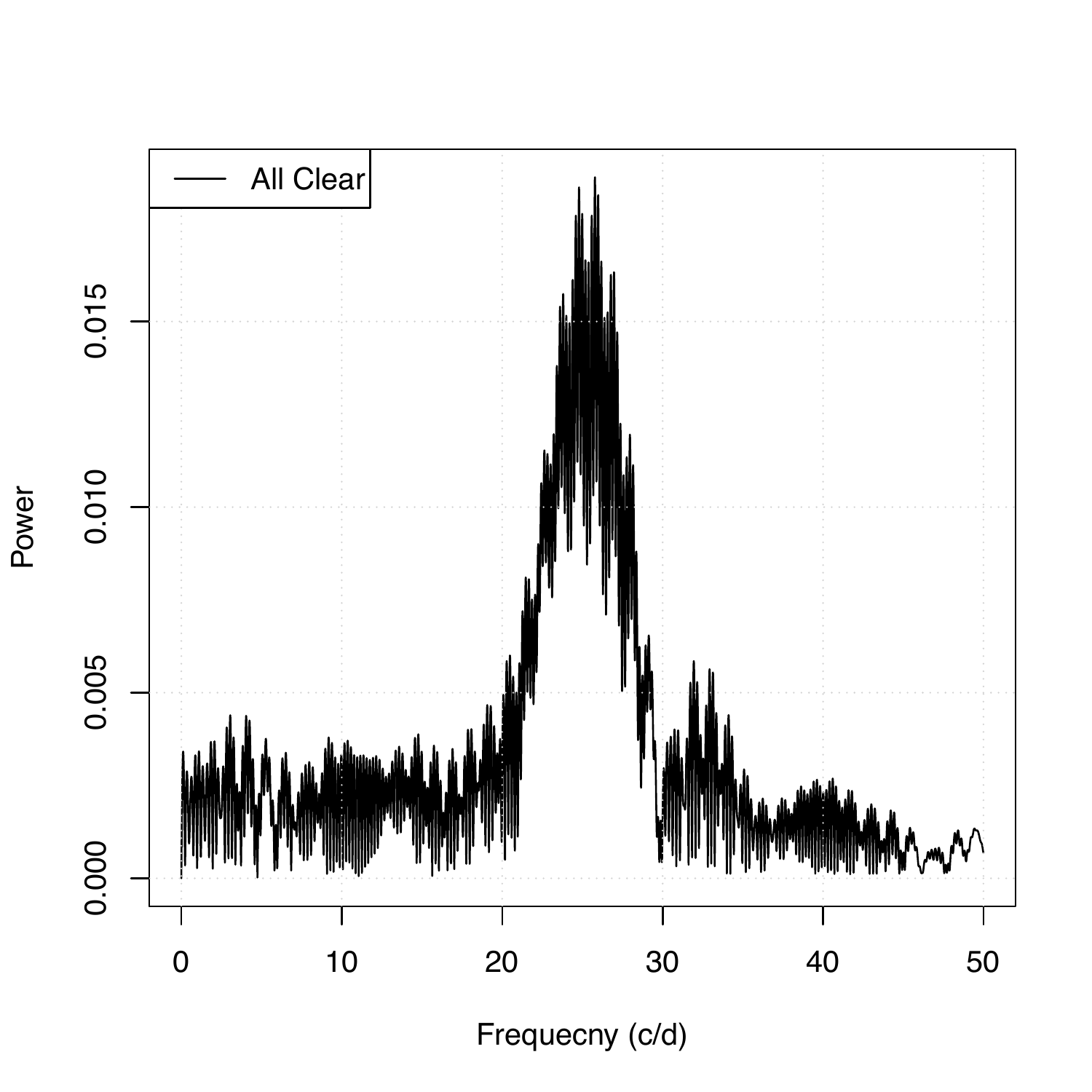}
  \includegraphics[width=3in, height=3in]{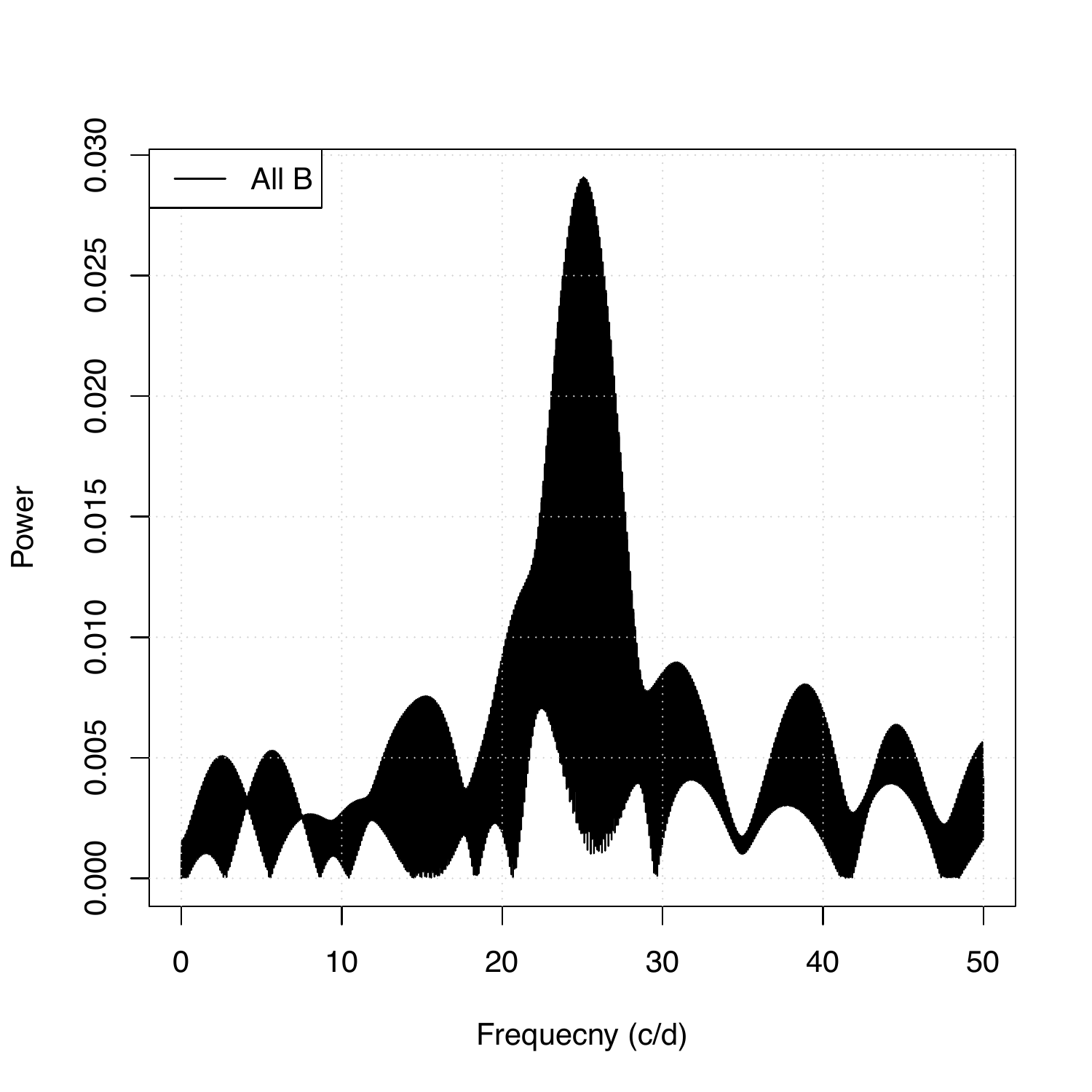}
  \includegraphics[width=3in, height=3in]{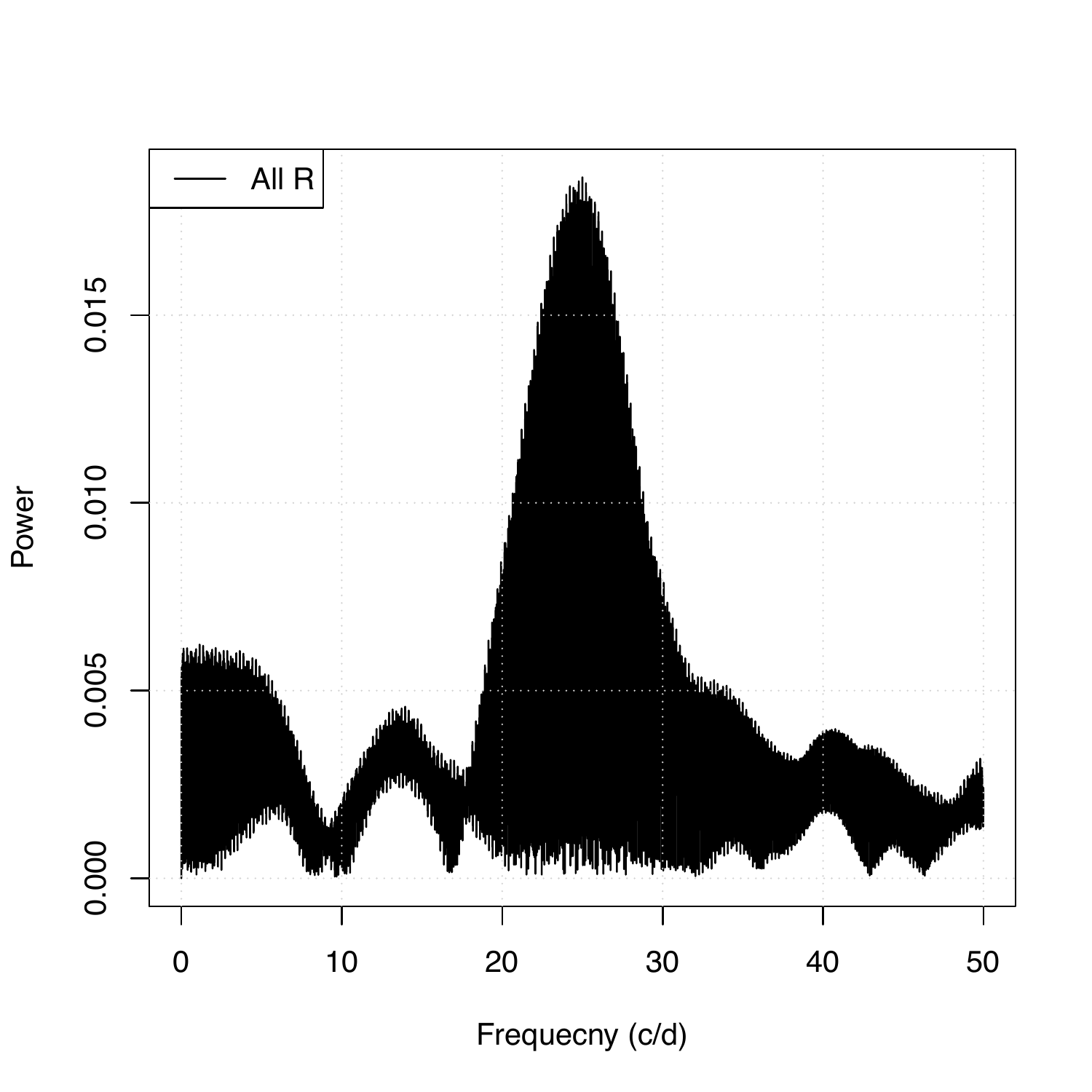}
  \includegraphics[width=3in, height=3in]{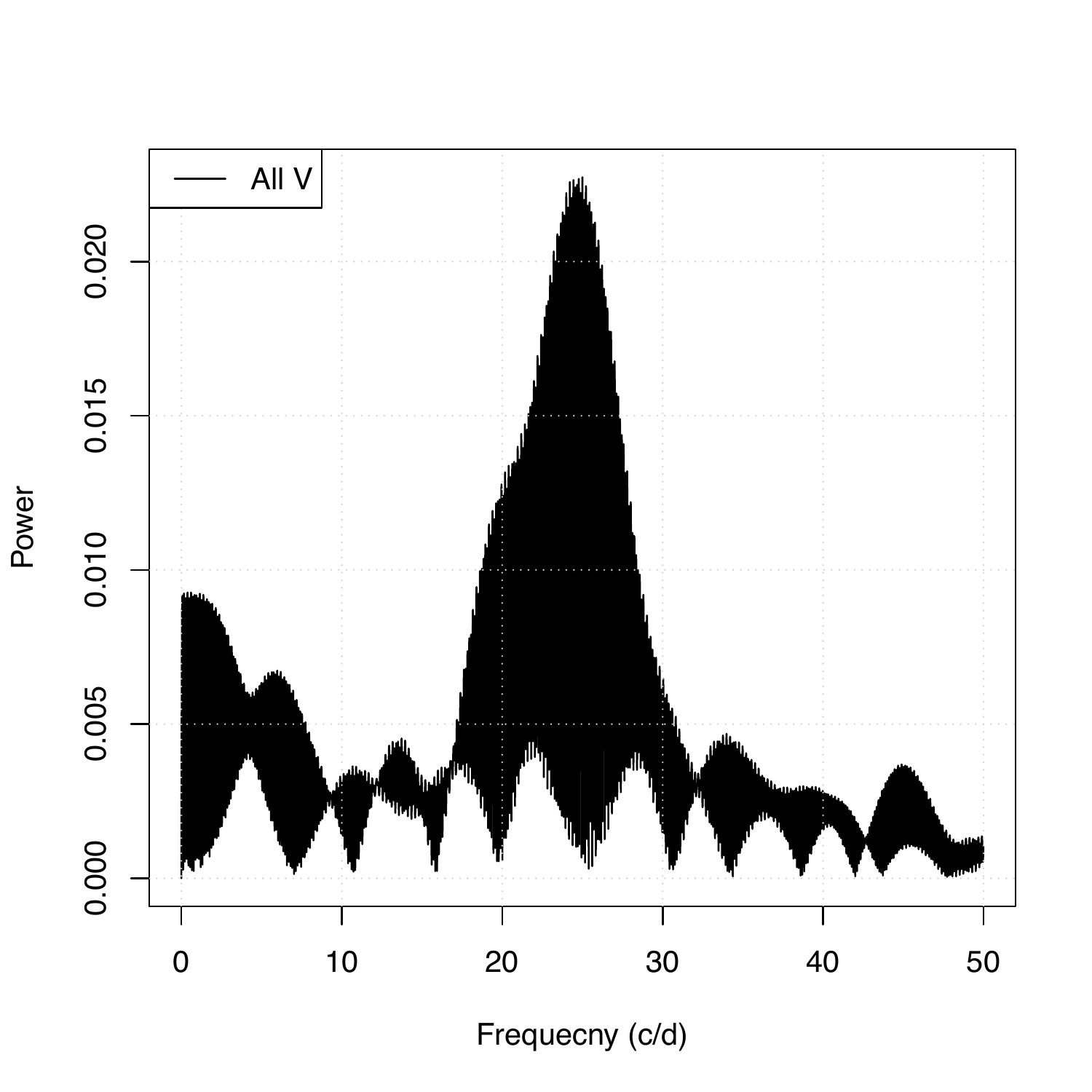}
  \caption{Power spectra for each filter. Plotted frequencies were obtained by combining all available data for the given filter.}
\label{powerspectra}
\end{figure*}

Table \ref{tabfreq} lists the obtained frequencies, associated errors and their corresponding periods in days. The error of each frequency were estimated using Monte-Carlo simulations as described in \cite{LenzBreger05}. Power spectra of the combined data of each filter are given in Fig. \ref{powerspectra}.

Figs. \ref{lc1}, \ref{lc2}, and \ref{lc3} shows light curves of the object in different filters (i.e. B, V, R, and Clear). In each panel, corresponding best-fit frequency given in Table \ref{tabfreq} is shown as the model (solid) line.

\begin{table}
\caption{List of detected principal frequencies ($f_{1}$), associated errors, signal-to-noise ratios, and the corresponding periods for the variability.}
\begin{center}
\renewcommand{\arraystretch}{1.4}
\setlength\tabcolsep{3pt}
\begin{tabular}{cccccc}
\hline
Date  &  Filter & Frequency	&Frequency Error& SNR    & Period 	\\ 
      &		    &  (c/d)	&(c/d) &            &(hours)     \\
\hline
20200223  & R     & 24.2793279 & 0.2941306 & 111.3         & 0.9884953     \\
20200225  & Clear & 25.6943480 & 0.1389742 & 13.1         & 0.9340576   \\
20200229  & Clear & 28.6745407 & 0.5177898 & 23.4        & 0.8369795    \\
20200301  & Clear & 25.2245375 & 0.0669629 & 27.4          & 0.9514546    \\
20200303  & R     & 25.0000000 & 0.2827345 & 16.8          & 0.9600000    \\
20200303  & V     & 25.1806881 & 0.1503901 & 67.1          & 0.9531113    \\
20200312  & R     & 24.7349823 & 0.2385561 & 77.3         & 0.9702858    \\
20200312  & V     & 24.3442751 & 0.1975498 & 66.4         & 0.9858588    \\
20201117  & B     & 24.7007410 & 0.2037593 & 32.0         & 0.9716308    \\
20201127  & B     & 25.0000000 & 0.1506077 & 23.1          & 0.9600000    \\
\hline
All Data  & Clear & 25.7830046 & 0.0022979 & 6.2          & 0.9308457    \\
All Data  & B     & 25.0692592 & 0.0001021 & 20.6         & 0.9573478    \\
All Data  & V     & 24.9956095 & 0.0016460 & 13.5         & 0.9601686    \\
All Data  & R     & 24.9994605 & 0.0272495 & 10.8          & 0.9600207   \\
\hline
\end{tabular}
\label{tabfreq}
\end{center}
\end{table}

For a first approach we assume all the periods with equal weights and obtained an average period for the variable as $P=0.940049$ hrs. The best observation night is the 20200301 (UT) based on the temporal coverage and the weather condition of that night. The average period that we obtained using all available data is consistent with the period obtained in the best night of $P=0.9514546$ hrs.

\begin{figure*}
\centering
  \includegraphics[width=3in, height=3in]{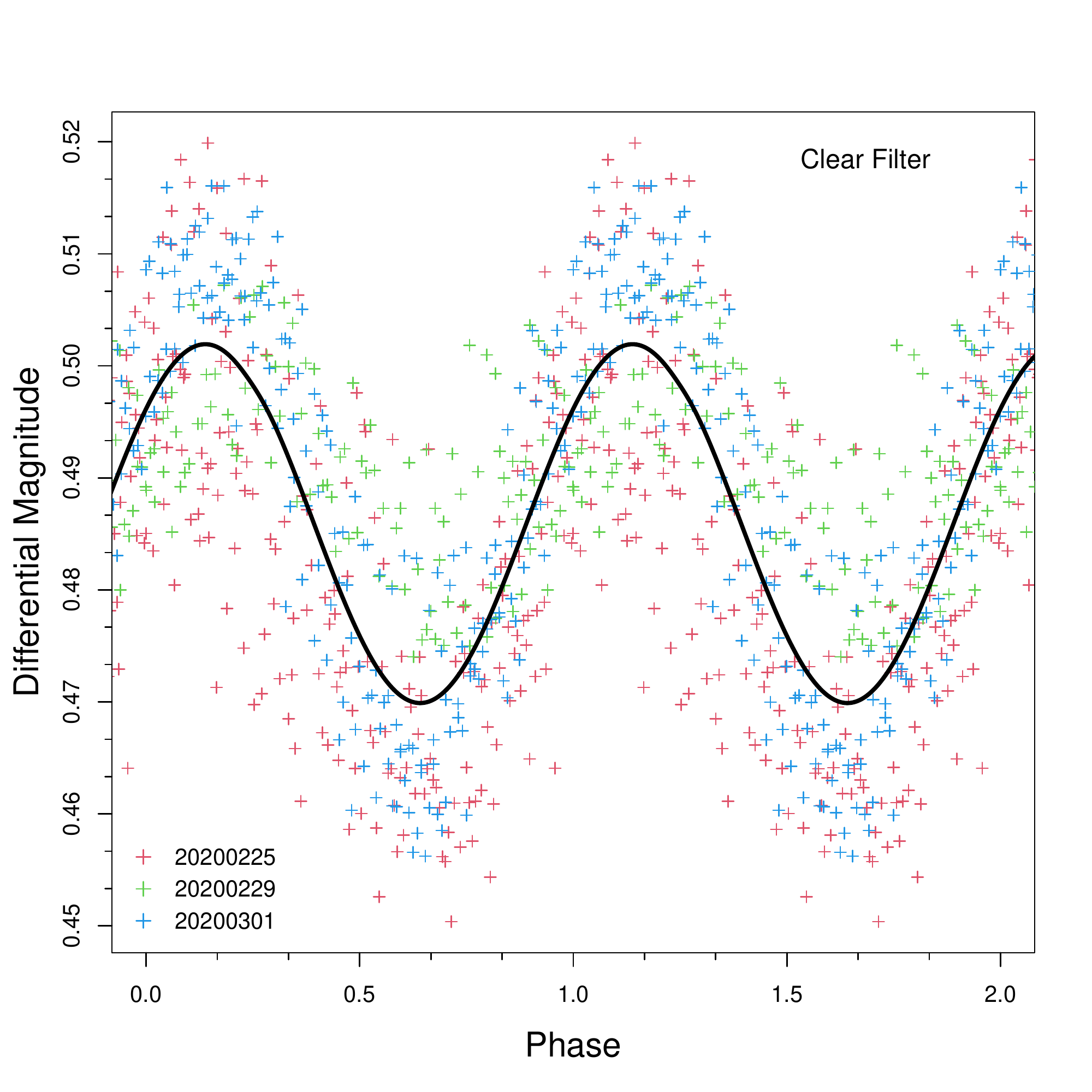}
  \includegraphics[width=3in, height=3in]{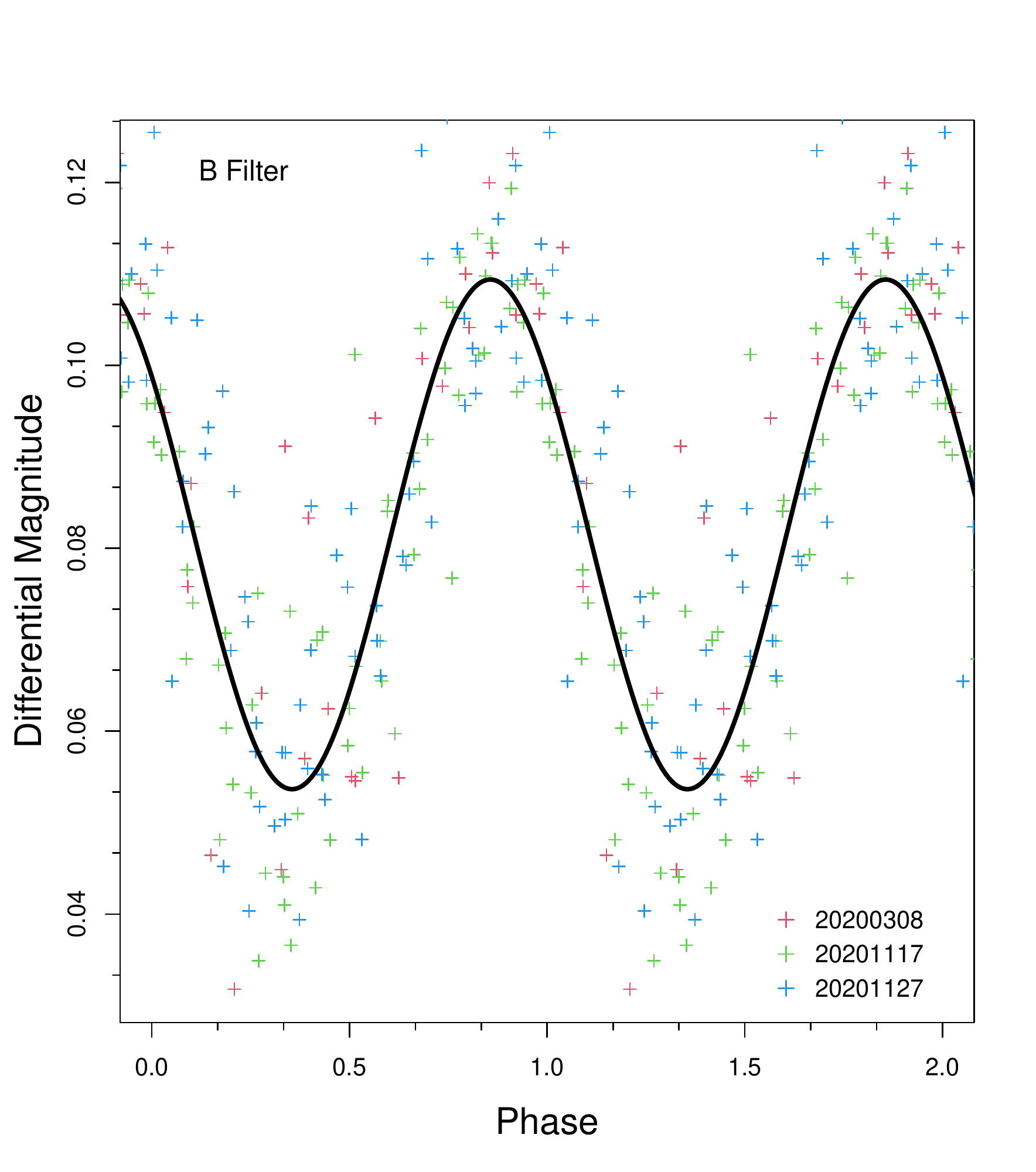}
  \includegraphics[width=3in, height=3in]{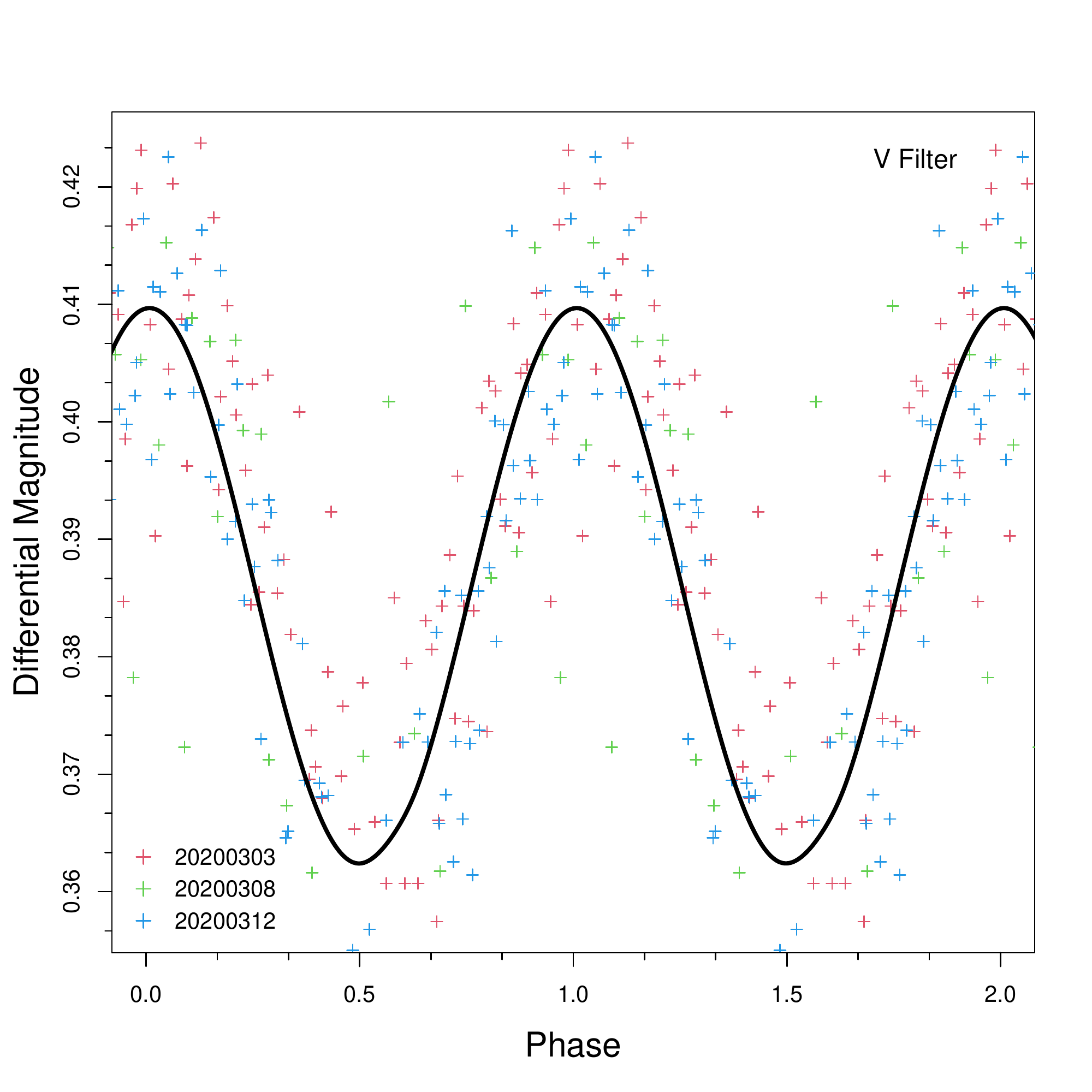}
  \includegraphics[width=3in, height=3in]{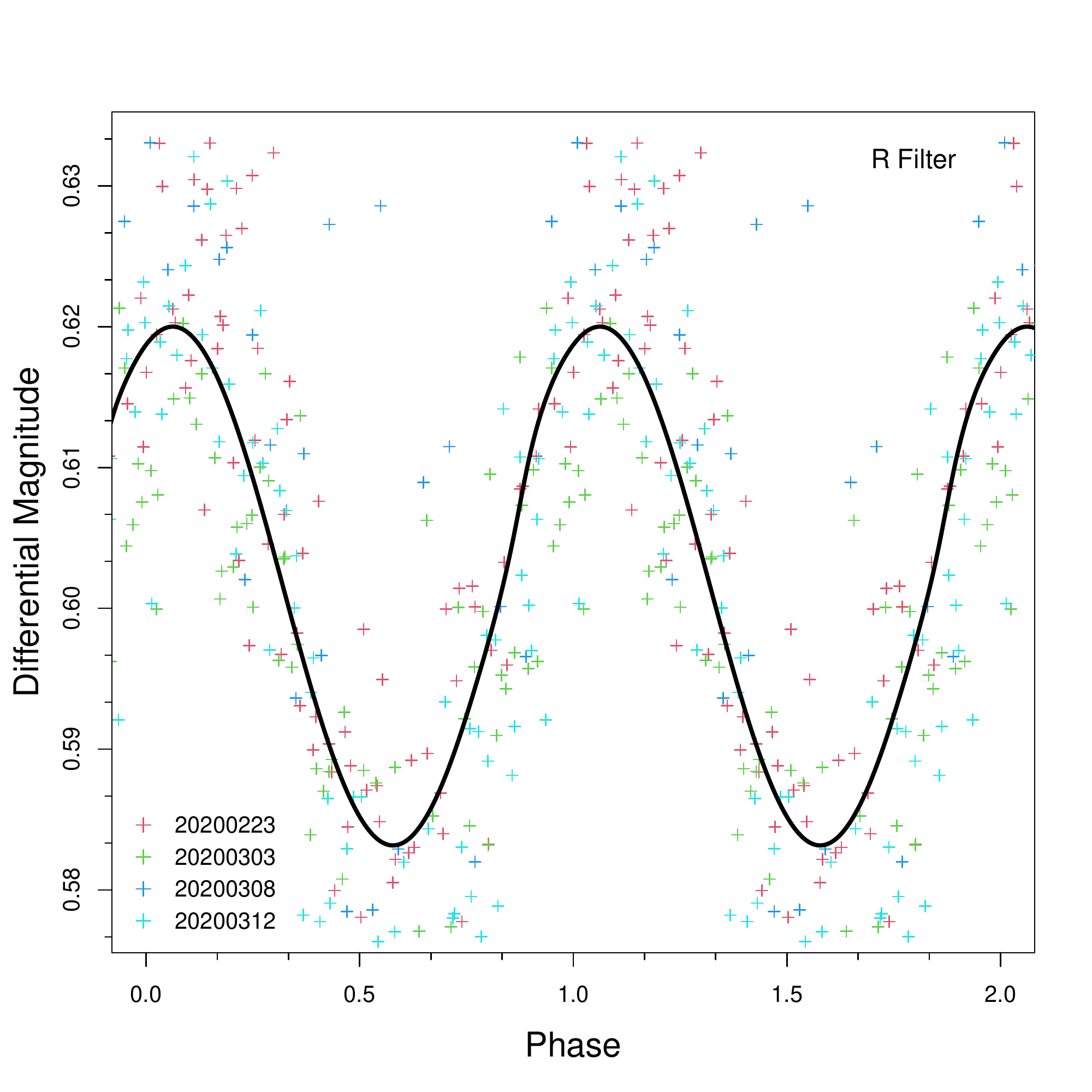}
  \caption{Phase diagram of the time-series data folded on the principal frequency obtained for each filter.}
\label{lcmodel}
\end{figure*}

\subsection{Times of Maxima}

We computed times of maxima of the variable using the Kwee \& van Woerden method \citep{KvW56}. This method requires a homogeneous temporal coverage of the maximum (or the minimum in eclipsing binaries). Thus, we use the data halfway from the maximum light at each side. In result, we omit maxima when the ascending or the descending portion of the light curve is not complete. In this way, we ensure the precision of the maxima times given in Table \ref{tabmaxima}. All times in the table are converted into Heliocentric Julian Date (HJD). Since we impose the completeness of the light curve for the times of maximum, we could also measure the amplitude of the light variation associated with a given maximum.

\begin{table}
\caption{List of maxima times. Table lists UT date, maximum time (HJD), uncertainty of the maximum (days), filter of the light curve, and the amplitude of the variation.}
\begin{center}
\renewcommand{\arraystretch}{1.4}
\setlength\tabcolsep{3pt}
\begin{tabular}{ccccc}
\hline
UT Date & $T_{max}$       & Uncertainty & Filter & Amplitude  \\         
(d.m.y) & (HJD 2400000+)  &    (days)    &        & (mag)    \\
\hline
23.02.2020 & 58903.34998 & 0.00119 & R & 0.03234\\
23.02.2020 & 58903.39220 & 0.00249 & R & 0.03182\\
25.02.2020 & 58905.25565 & 0.00080 & Clear & 0.05644\\
25.02.2020 & 58905.35724 & 0.00028 & Clear & 0.02894\\
25.02.2020 & 58905.40670 & 0.00068 & Clear & 0.03300\\
29.02.2020 & 58909.40730 & 0.00052 & Clear & 0.02837\\
29.02.2020 & 58909.44399 & 0.00100 & Clear & 0.02349\\
29.02.2020 & 58909.47835 & 0.00119 & Clear & 0.01529\\
01.03.2020 & 58910.29372 & 0.00068 & Clear & 0.04490\\
01.03.2020 & 58910.33481 & 0.00045 & Clear & 0.04989\\
01.03.2020 & 58910.37437 & 0.00059 & Clear & 0.04454\\
01.03.2020 & 58910.41013 & 0.00077 & Clear & 0.03370\\
03.03.2020 & 58912.38964 & 0.00089 & R & 0.03333\\
03.03.2020 & 58912.42990 & 0.00064 & R & 0.03721\\
03.03.2020 & 58912.47129 & 0.00184 & R & 0.00410\\
03.03.2020 & 58912.27010 & 0.00097 & V & 0.02332\\
03.03.2020 & 58912.38850 & 0.00134 & V & 0.04305\\
03.03.2020 & 58912.43070 & 0.00027 & V & 0.05308\\
03.03.2020 & 58912.47219 & 0.00052 & V & 0.02609\\
08.03.2020 & 58917.38538 & 0.00072 & R & 0.03565\\
08.03.2020 & 58917.38557 & 0.00066 & V & 0.03391\\
08.03.2020 & 58917.38625 & 0.00105 & B & 0.05069\\
12.03.2020 & 58921.30899 & 0.00089 & R & 0.04743\\
12.03.2020 & 58921.34873 & 0.00127 & R & 0.04727\\
12.03.2020 & 58921.39172 & 0.00015 & R & 0.04677\\
12.03.2020 & 58921.30995 & 0.00136 & V & 0.05819\\
12.03.2020 & 58921.35010 & 0.00075 & V & 0.06229\\
12.03.2020 & 58921.39137 & 0.00123 & V & 0.06299\\
17.11.2020 & 59171.40134 & 0.00046 & B & 0.02148  \\
17.11.2020 & 59171.44284 & 0.00023 & B & 0.01820 \\
17.11.2020 & 59171.48147 & 0.00011 & B & 0.01985 \\
27.11.2020 & 59181.33463 & 0.00131 & B & 0.02048 \\
\hline
\end{tabular}
\label{tabmaxima}
\end{center}
\end{table}

We computed the ephemeris of the variable by using the all available maximum times irrespective of the filter used. Employing a linear fit to the maximum times reveals the ephemeris below:

\begin{equation}
    T_{max} \ (HJD) = 0.039439 \times E + 58903.36375
\end{equation}

In Fig. \ref{lcmodel} we folded all observations according to the period obtained from the data of the same filter. Each panel in the figure shows specific filters (i.e. V, R, and Clear) and different colors represent different observing nights.

\section{Variability Type}

Variability behaviour of the source is similar with those of short period pulsators. The period of variability and the shape of the light curve imply that the variable is likely a Delta Scuti star \citep{Sterken05}. However, we wanted to test this with a machine-learning approach described below.

To determine the variability type of the object, we used UPSILoN (AUtomated Classification of Periodic Variable Stars using MachIne LearNing) package from the Python library \citep{Kim2016}. UPSILoN essentially aims to automatically classify variable sources from their optical light curve. It essentially contains two parts; the first part is extracting features from a light curve, and the other part is utilizing features for classification.

The UPSILoN is using the random forest model for classification. Random forest classifiers are based on a collection of decision trees \citep{Huang2020,Quinlan1993}. For each decision tree, a random subset of input features is selected and used to build the tree. Models are using 16 extracted features to predict a class.

The package is able to differentiate variable sources of six variability types. These types are the followings: Delta Scuti, RR Lyrae, Cepheid, Type-II Cepheid, eclipsing binary, and long-period variables. The algorithm is trained with light curves obtained from OGLE and EROS-2 surveys. A total of 143,923 stars have been used for training sample where 3209 of them were of type Delta Scuti.

We compiled all time series data according to the observed filter and fed into the UPSILoN. For all data the most plausible classification is returned as Delta Scuti. The probability of being a Delta Scuti variable for the different filters are as follows: 69\% (B), 83\% (V), 76\% (R), and 82\% (Clear). In conclusion, it is possible to state the new variable is of Delta Scuti with a mean probability of 78\%.

We should mention that this detection of the variability type does not include all known variable classes but the well-known ones. In order to reach a conclusion on the variability type, one needs to have the luminosity and the effective temperature of the object as we show in the next section.

\subsection{Physical parameters}

Following the determination of the variability type as Delta Scuti, we computed the absolute magnitude ($M_{v}$) and the intrinsic B-V color index using the calibrations given by \cite{McNamara11}:

\begin{equation}
\label{eq_PL}
M_{v} = (-2.89 \pm 0.13) \ log(P) - (1.31 \pm 0.10)
\end{equation}

\begin{equation}
\label{eq_BV}
(B-V)_{0} = (0.125 \pm 0.006) \ log(P) + (0.397 \pm 0.006)
\end{equation}

\noindent Eq. \ref{eq_PL} is actually the period-luminosity relation. P is the pulsation period in days in the both Eqs. \ref{eq_PL} and \ref{eq_BV}. Both equations are valid for solar metallicity but this is not a concern as Delta Scuti variables are mostly considered as solar-type stars. Thus, we obtained the absolute magnitude in V-band as $M_{v}=2.76\pm0.28$ and intrinsic color index of $(B-V)_{0}=0.22\pm0.01$. For a sanity check, we also determined the $(B-V)$ of the variable using the transformation equations given in \cite{Karaali2005} together with the Sloan magnitudes given in Table \ref{infotab} and found exactly the same value of $(B-V)_{0}=0.22$.

This $(B-V)_{0}$ translates into an effective temperature of $7770\pm200$ K when we take into account the calibrations given by \cite{Cox2000}. This color index also corresponds to the spectral type A7 \citep{Bessell1979}. The error value of the $\mathrm{T_{eff}}$ mainly comes from the color index calibrations as stated in \cite{Bessell1979} and \cite{Sekiguchi2000}. However, an exact error term for the effective temperature was not given in those calibrations. Thus, we adopted here a relatively conservative error (i.e. 200 K) for the effective temperature. The error contribution of the Eq. 3 is very small compared to this error term.

This effective temperature and the spectral type is consistent with typical properties of Delta Scuti stars \citep{Breger2007}.

The variable has been observed with Gaia \citep{Gaia2016} (\textit{Gaia EDR3 982385743606251648}) and we obtained the parallax from the Gaia EDR3 catalogue \citep{GaiaEDR3} as 0.5627 mas ($\pm 0.0178$ mas). Therefore, the distance to the variable based on its Gaia parallax is $1722_{-58}^{+54}$ pc.

Adopting a bolometric correction of $BC=-0.122$ \citep{Cox2000} yields $M_{bol}=2.64\pm0.28$ which then used together with $M_{bol,\odot}=4.74$ to derive the luminosity of the variable as $L=6.93_{-2.04}^{+1.58} \ L_{\odot}$.

\begin{figure}
\centering
  \includegraphics[width=\columnwidth]{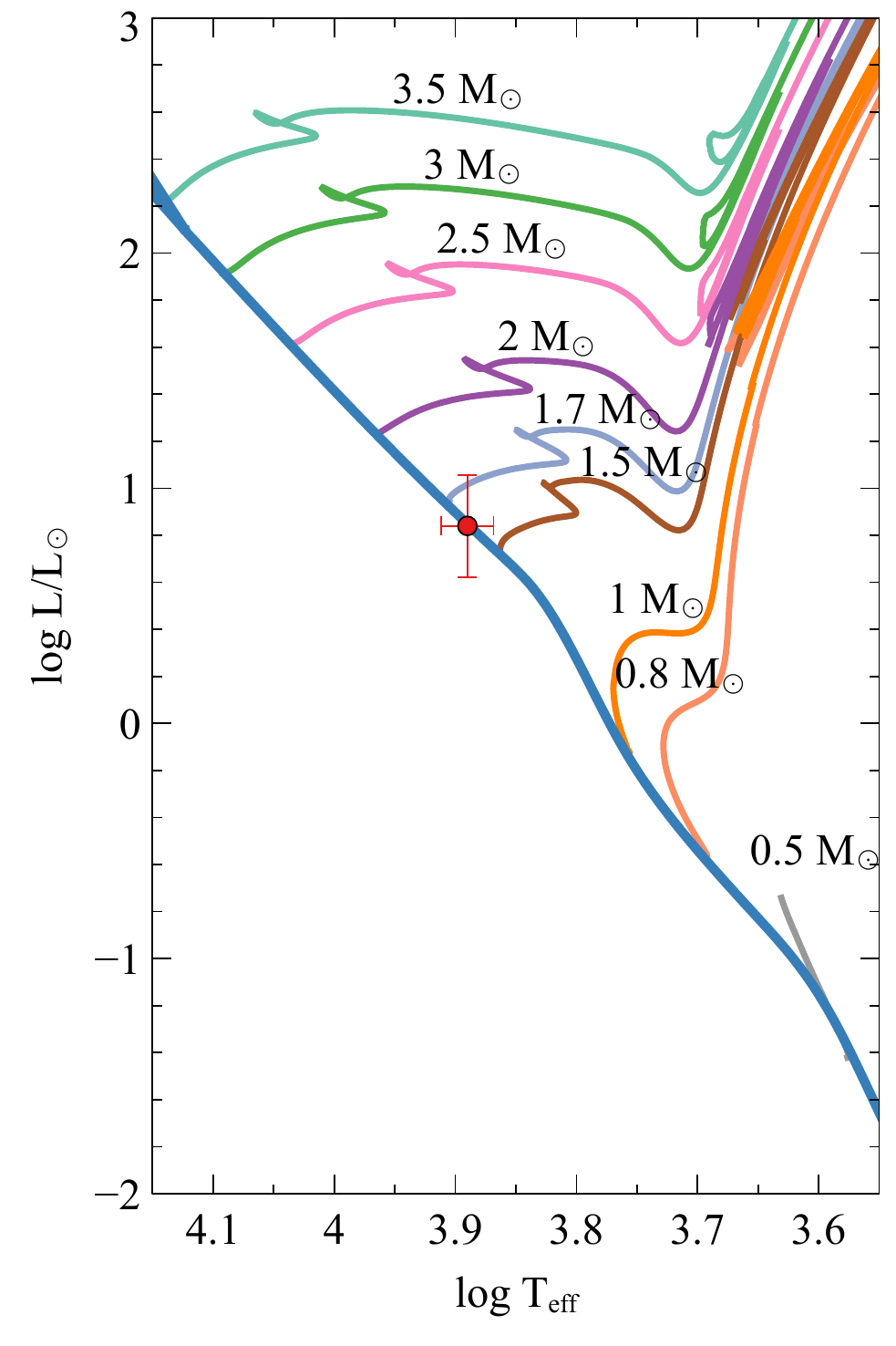}
  \caption{Evolutionary tracks for low-mass stars computed with MESA codes \protect\citep{Paxton2011}. The location of the variable is marked with a red point which is between the tracks of $1.5-1.7 \ M_{\odot}$ stars. Error on the luminosity comes from the period-luminosity relation and the bolometric correction whereas the error on the effective temperature mainly comes from the color index calibration given in \protect\cite{Cox2000}.}
\label{evol}
\end{figure}

The derived values of effective temperature and luminosity locate the variable between the evolutionary tracks of $1.5-1.7 \ M_{\odot}$ stars as shown in Fig. \ref{evol}. The evolutionary tracks shown in Fig. \ref{evol} were computed using MESA (Modules for Experiments in Stellar Astrophysics) codes \citep{Paxton2011}. The bolometric magnitude mass calibration given in \cite{Jafarzadeh17} reveals the mass for the variable as $M=1.57\pm0.1 \ M_{\odot}$ which is consistent with the evolutionary tracks. This result is consistent with the location of Delta Scuti stars on the H-R diagram.

\section{Summary and Conclusions}

We present the detection of a new variable in the field of exoplanet host star XO-2. The new variable has been observed in 10 nights with different standard photometric filters besides white light observations. Both the shape of the light curves and the automated classification based on machine learning algorithms reveal the candidate is a Delta Scuti with latter have the probability of 78\%.

Examining individual light curves allowed us to measure 32 times of maxima which also enabled to determine the ephemeris of the variable for the first time. Hence, the pulsation period of the new variable determined as $P=0.039439$ days.

With the help of the fundamental calibrations we could obtain the astrophysical parameters of the variable and located it on the evolutionary tracks at the H-R diagram. Thus, the new variable has a mass between $1.5-1.7 \ M_{\odot}$. The variable is of A7 spectral type with an approximate effective temperature of 7725 K which are consistent with typical properties of Delta Scuti stars.

Further observations are needed to investigate any possible variability of amplitude and/or frequency changes in the system or to study multi-mode pulsation behaviour in detail. Recent observations of the TESS in the Sector 47 cover the variable where only the full-frame images are available at the moment. Additional ground-based observations can be coupled together the TESS short cadence observations to perform a deeper study of the new variable.

\section*{Acknowledgements}
IST40 is one of the observational facility of the Istanbul University Observatory. This study was funded by Scientific Research Projects Coordination Unit of Istanbul University with project numbers: BAP-3685 and FBG-2017-23943. SA and FKY acknowledge the support from TUBITAK through the ARDEB-1001 program with the project number 118F042. Authors are thankful to Ahmet Dervişoğlu for his help on the stellar evolutionary tracks.

This work has made use of data from the European Space Agency (ESA) mission
{\it Gaia} (\url{https://www.cosmos.esa.int/gaia}), processed by the {\it Gaia}
Data Processing and Analysis Consortium (DPAC,
\url{https://www.cosmos.esa.int/web/gaia/dpac/consortium}). Funding for the DPAC
has been provided by national institutions, in particular the institutions
participating in the {\it Gaia} Multilateral Agreement.

\label{refer}

\bibliographystyle{tjaa}
\bibliography{IST40spp}

\bsp	
\label{lastpage}
\end{document}